\documentclass[letterpaper,twocolumn,10pt]{article}
\usepackage{usenix}

\usepackage{tikz}
\usepackage{amsmath}
\usepackage{svg}
\usepackage{filecontents}

\usepackage{booktabs}
\usepackage{multirow}
\usepackage{array}
\usepackage{makecell}
\usepackage{graphicx}   
\usepackage{ragged2e}   

\newcolumntype{L}[1]{>{\RaggedRight\arraybackslash}m{#1}}
\newcolumntype{C}[1]{>{\centering\arraybackslash}m{#1}}

\usepackage{xcolor}
\definecolor{mygray}{gray}{0.9}

\usepackage{booktabs}
\usepackage{multirow}
\usepackage{array}
\usepackage{makecell}
\usepackage{graphicx}
\usepackage{ragged2e}
\usepackage{adjustbox}

\newcolumntype{L}[1]{>{\RaggedRight\arraybackslash}m{#1}}
\newcolumntype{C}[1]{>{\centering\arraybackslash}m{#1}}

\usepackage{graphicx}
\usepackage{makecell}

\newcommand{\symcircle}{0.65ex}

\newcommand{\yes}{%
  \tikz[baseline=-0.55ex]\fill (0,0) circle (\symcircle);%
  \xspace
}

\newcommand{\partmark}{%
  \tikz[baseline=-0.55ex]{%
    \draw[line width=0.35pt] (0,0) circle (\symcircle);
    \begin{scope}
      \clip (0,0) circle (\symcircle);
      \fill (-\symcircle,-\symcircle) rectangle (0,\symcircle);
    \end{scope}
  }%
  \xspace
}

\newcommand{\quasi}{%
  \tikz[baseline=-0.55ex]\draw[line width=0.35pt] (0,0) circle (\symcircle);%
  \xspace
}

\newcommand{\no}{$-$\xspace}

\newcommand{\leftrule}[1]{
    \multicolumn{1}{|c}{#1}
}

\newcommand{\verttext}[1]{%
  \rotatebox{70}{\footnotesize #1}%
}

\usepackage{booktabs}
\usepackage{multicol}
\usepackage{multirow}
\usepackage{fontawesome}
\usepackage{array}
\usepackage{xcolor}
\usepackage{mathtools}
\usepackage{colortbl}
\usepackage{xspace}
\usepackage{mathtools}
\usepackage{kmath}

\usepackage{booktabs}
\usepackage{multicol}
\usepackage{multirow}
\usepackage{fontawesome}
\usepackage{array}
\usepackage{xcolor}
\usepackage{mathtools}
\usepackage{colortbl}
\usepackage{xspace}
\usepackage{mathtools}
\usepackage{kmath}

\usepackage{enumitem}
\usepackage{graphicx}
\usepackage{colortbl}
\usepackage{array}
\usepackage[table,xcdraw]{xcolor}
\usepackage{multicol}
\usepackage{multirow}
\usepackage{booktabs}
\usepackage{arydshln}
\usepackage{ragged2e}
\usepackage{subcaption}
\usepackage{url}
\usepackage{amsmath}
\usepackage{mathtools}
\usepackage{siunitx}
\usepackage{longtable}
\usepackage[most]{tcolorbox}
\usepackage{todonotes}
\usepackage{balance}
\usepackage[most]{tcolorbox}

\usepackage{booktabs}
\usepackage{tabularx}
\usepackage{multirow}
\usepackage{array}

\newcolumntype{L}[1]{>{\raggedright\arraybackslash}p{#1}}

\usepackage{booktabs}
\usepackage{graphicx} 
\usepackage{array}

\usepackage{xurl} 

\newtcolorbox{rqsummarybox}[2][]{
  colback=gray!5,
  colbacktitle=gray!20,
  colframe=black!50,
  coltitle=black,
  boxrule=0.6pt,
  titlerule=0pt, 
  title={Summary #2},
  fonttitle=\bfseries,
  boxsep=1ex,
  left=1ex,
  right=1ex,
  top=0.8ex,
  bottom=0.8ex,
  breakable,
  #1
}

\makeatletter
\renewenvironment{quote}
  {\list{}{\leftmargin=1\parindent
           \rightmargin=1\parindent
           \topsep=0pt
           \partopsep=0pt
           \parsep=0pt
           \itemsep=0pt}%
   \item\relax\itshape}
  {\endlist}
\makeatother

\newcommand{\paragraphb}[1]{\vspace{0.04in}\noindent{\bf #1.} }

\begin{document}

\date{}

\title{\Large \bf Understanding U.S. Users' Security and Privacy Transparency Needs for Consumer-Facing Generative AI}

\def\plainauthor{Jiaxun Cao, Yu Dong, Chunxi Zhan, Rithvik Neti, Sai Teja Peddinti, Pardis Emami-Naeini}
\author{
    Jiaxun Cao$^{\dagger}$,
    Yu Dong$^{\ddagger*}$,
    Chunxi Zhan$^{\ddagger*}$,
    Rithvik Neti$^{\dagger}$,
    Sai Teja Peddinti$^{\mathsection}$,
    Pardis Emami-Naeini$^{\dagger}$ \\
    \\
    \textit{$^{\dagger}$Duke University}, 
    \textit{$^{\ddagger}$Duke Kunshan University}, 
    \textit{$^{\mathsection}$Google} \\
}

\maketitle
\thecopyright
\begingroup
\renewcommand\thefootnote{*}
\footnotetext{The two authors contributed equally to this work.}
\endgroup

\begin{abstract}
    \label{sec:abstract}
Users increasingly rely on consumer-facing generative AI (GenAI) for tasks ranging from everyday needs to sensitive use cases. Yet, it remains unclear whether and how existing security and privacy (S\&P) communications in GenAI tools shape users’ adoption decisions and experiences. Understanding how users seek, interpret, and evaluate S\&P information is critical for designing usable transparency that users can \emph{trust} and \emph{act on}. We conducted semi-structured interviews and design sessions with 21 U.S. GenAI users. Our findings suggest that available S\&P information rarely drove initial adoption in practice, as participants often perceived it as incomplete, ineffective, or not credible. Instead, they relied on rough proxies (e.g., popularity) to infer S\&P practices. After adoption, S\&P uncertainty constrained participants’ willingness to use GenAI tools, especially for high-stakes purposes, and, in some cases, contributed to discontinued use. Participants therefore called for transparency that supports decisions and actions through trustworthy information (e.g., independent evaluations) and usable interfaces (e.g., on-demand disclosure). We categorize participants’ desired design practices into five dimensions to facilitate systematic future investigation into best practices. We conclude with recommendations for researchers, designers, and policymakers to improve S\&P transparency in consumer-facing GenAI.
\end{abstract}
\section{Introduction} \label{sec:intro}
Generative AI (GenAI) has seen widespread adoption for diverse purposes, from everyday information seeking to highly sensitive, personal use cases such as health, legal, and financial consultations~\cite{schneiders2025objection, zhang2024s}. Security and privacy (S\&P) researchers have warned about multiple GenAI risks, including adversarial attacks, data breaches, and the sharing or sale of personal data~\cite{liu2023prompt, zhang2024s, kshetri2023cybercrime, carlini2022quantifying, carlini2021extracting, zhang2023counterfactual}. Yet, empirical evidence suggests that these S\&P intrusions often occur without users' awareness or consent~\cite{lee2024deepfakes}. Even when users are aware, their mental models of GenAI data practices are often inaccurate~\cite{kwesi2025exploring, ma2025privacy, malki2025hoovered, zhang2024s, ali2025understanding}. Prior work shows that users often overestimate the protections provided by GenAI companies~\cite{malki2025hoovered, kwesi2025exploring, ma2025privacy}. For instance, mistakenly assuming that ChatGPT complies with the Health Insurance Portability and Accountability Act (HIPAA) when used for mental health~\cite{kwesi2025exploring}, or that OpenAI is responsible for privacy protections for third-party generative pre-trained transformers (GPTs)~\cite{ma2025privacy}.

These prevalent S\&P misconceptions point to a gap in current GenAI transparency -- users are likely receiving limited or ineffective communication about data practices and associated risks. Much of the current S\&P transparency work around GenAI is oriented toward expert audiences, presuming substantially higher technical literacy than most end users. Examples include Google Model Cards~\cite{mitchell2019model}, Meta System Cards~\cite{meta_system_cards}, and IBM AI FactSheets~\cite{ibm_ai_factsheets_docs}. Policy efforts in the U.S. (e.g., Colorado SB24-205~\cite{colorado_sb24_205_2024}) and beyond (e.g., the EU AI Act~\cite{eu_ai_act_2024}) have also remained limited, tending to emphasize organizational accountability and overlook requirements for consumer-facing S\&P transparency~\cite{nist_ai_rmf_2023, nist_genai_profile_2024}.

However, S\&P transparency can shape consumers' adoption decisions and post-adoption experiences~\cite{egelman2009timing, tsai2011effect,murmann2021privacy_notifications, emami2021privacy}. When transparency is reliable, actionable, and presented in forms that users can readily interpret and act on, it can support informed S\&P decision-making and help build trust in the technology~\cite{schaub2015design,murmann2021privacy_notifications}. For instance, prior work has proposed privacy labels that present salient information in a digestible format and impact users' willingness to purchase Internet of Things (IoT) devices~\cite{kelley2009nutrition, emami2021informative, emami2023consumers}. In the GenAI context, however, comparable empirical evidence remains limited -- it is unclear whether, what, and how S\&P information influences users' adoption decisions and ongoing use of GenAI tools.

Beyond S\&P transparency content, prior work has articulated a broader design space spanning multiple dimensions~\cite{schaub2015design}, including timing~\cite{egelman2009timing, good2007noticing}, channel~\cite{schaub2015design}, modality~\cite{wogalter2002based}, and user control~\cite{schwartz2004paradox}. Each technology introduces distinct challenges and opportunities across these dimensions that must be accounted for when designing effective transparency~\cite{schaub2015design}. Motivated by this perspective, we examine whether and how users currently seek, understand, and evaluate existing GenAI S\&P information, and what improvements they hope to see in the content and design of GenAI S\&P transparency. We ask three research questions (RQs):

\begin{itemize}[itemsep=0pt, topsep=0pt, parsep=0pt, partopsep=0pt]
    \item \textbf{RQ1:} What, if any, S\&P information about consumer-facing GenAI tools factors into users' adoption decisions and post-adoption experiences?

    \item \textbf{RQ2:} What are users’ current practices and challenges in seeking, understanding, and evaluating S\&P information about consumer-facing GenAI tools?

    \item  \textbf{RQ3:} What are users’ preferences and expectations for S\&P transparency in consumer-facing GenAI tools?
\end{itemize}

To answer these questions, we conducted 21 semi-structured interviews followed by design sessions. Our findings suggest that existing S\&P information is often perceived as incomplete, ineffective, and lacking credibility, and rarely drives GenAI adoption decisions. At the same time, participants described needs for effective S\&P communication in GenAI -- needs they expected to become increasingly consequential as their high-stakes use grows. Across interviews and design sessions, participants articulated specific information needs (e.g., independent evaluations) and proposed a rich set of design practices across five dimensions. Grounded in these findings, we discuss design opportunities for advancing GenAI S\&P transparency and offer actionable recommendations for S\&P researchers, designers, and policymakers.

As an exploratory study intended to spur future research on the emerging design space of GenAI S\&P transparency, we make three contributions: (1) empirical accounts of how users currently seek, understand, and evaluate GenAI S\&P information; (2) a synthesis of participant-informed transparency design practices into a set of dimensions for future design and evaluation; and (3) actionable implications and recommendations for S\&P researchers, designers, and policymakers to advance consumer-facing S\&P transparency in GenAI.
\section{Related Work} \label{sec:background}
\paragraphb{GenAI S\&P risks and users' (mis)conceptions} The S\&P risks of GenAI have been extensively investigated since its emergence. From a security perspective, GenAI systems are vulnerable to adversarial threats, including data poisoning~\cite{marchal2024generative}, prompt injection~\cite{liu2023prompt}, and spoofing~\cite{tao2023opening}. GenAI can also enable malicious use, such as generating convincing phishing messages~\cite{bird2023typology}. Privacy risks center on extensive data collection, data breaches, and inappropriate use or sale of personal data~\cite{ma2025privacy, zhang2024s, kshetri2023cybercrime, carlini2022quantifying, carlini2021extracting, zhang2023counterfactual}. With the increasingly human-like and intelligent nature of GenAI tools, users share highly sensitive data~\cite{zhang2024s, ma2025privacy, weidinger2022taxonomy, van2019chatbot}, such as medical records, personal trauma, and payslips~\cite{zhang2024s}. Such data can enable profiling, as GenAI systems may infer additional attributes, including sexual orientations, genders, and religious beliefs~\cite{ma2025privacy, weidinger2022taxonomy}. These sensitive data are often shared beyond GenAI companies, including with third parties that build applications on top of host GenAI models and may not guarantee users comparable privacy protections~\cite{ma2025privacy}.

Yet, these S\&P harms often unfold with limited user awareness and without informed consent~\cite{lee2024deepfakes, zhang2024s}. Extensive research shows that users’ mental models of GenAI data practices, such as how data is collected, retained, used for training, and reused, are often incomplete or incorrect~\cite{zhang2024s, ali2025understanding, ma2025privacy, malki2025hoovered, kwesi2025exploring, wang2025mental, zhang2025understanding}. For example, users may not recognize all stakeholders who can access their data, including internal staff who review conversations for safety~\cite{ali2025understanding, zhang2024s}. Prior work also finds that users misunderstand the scope of data controls and policies. For instance, Malki et al.~\cite{malki2025hoovered} report that users often assume deletion or training opt-out removes what the model has already learned. In practice, machine unlearning remains an unsolved problem~\cite{wang2024machine, xu2024machine, zhang2023review}. In sensitive contexts, such as mental health, users often assume that ChatGPT complies with HIPAA, even though it does not~\cite{kwesi2025exploring}.

Such extensive empirical evidence on users' misconceptions about GenAI S\&P practices reveals a significant gap in the current infrastructure for GenAI S\&P transparency. Specifically, existing S\&P information and how it is communicated may not sufficiently support users in understanding, trusting, and acting on data practices. Our work thus contributes by surfacing whether and what S\&P transparency needs users have for a more trustworthy GenAI ecosystem.

\begin{figure*}
    \centering
    \includegraphics[width=1\linewidth]{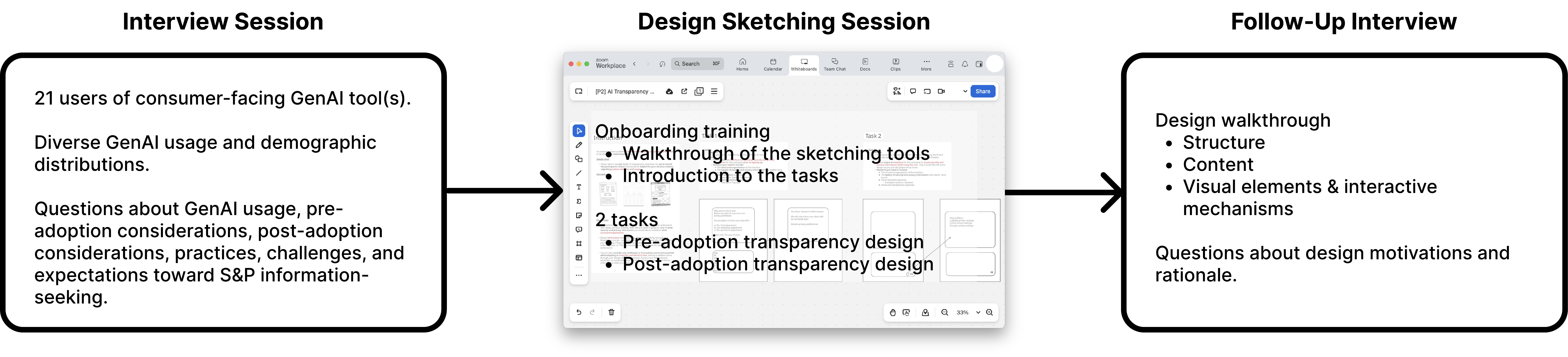}
    \vspace{-0.2in}
    \caption{Methodology overview of our main study. We conducted semi-structured interviews, followed by design sketching sessions and follow-up questions.}
    \label{fig:zoom}
\end{figure*}

\paragraphb{Existing S\&P transparency approaches} While GenAI S\&P transparency remains underexplored, existing research on S\&P transparency has examined a range of prior technologies~\cite{schaub2015design, emami2020ask, balebako2014your, cranor2012necessary, jensen2004privacy, fu2014field, cranor2006user, kelley2010standardizing, kelley2013privacy, langheinrich2002privacy, harkous2016pribots, freiberger2026helping}, including mobile apps~\cite{sunyaev2015availability, zimmeck2017automated}, IoT~\cite{emami2020ask, castelluccia2018enhancing}, and ubiquitous computing systems~\cite{langheinrich2002privacy}. A common effort to promote S\&P transparency is the use of privacy policies, yet they have been widely criticized for being lengthy and difficult to understand~\cite{jensen2004privacy, mcdonald2008cost}. Towards more effective S\&P communication, prior work highlights several design dimensions~\cite{schaub2015design}, including timing (when information is presented), channel (where it is presented), modality (how it is presented), and user control (how users can act on it).

To improve the readability of privacy policies, alternative approaches have emerged, such as privacy nutrition labels -- analogous to food nutrition labels that surface salient information in a concise, standardized format~\cite{kelley2009nutrition, emami2020ask, caven2025usability}. Prior work has identified and evaluated label factors that matter to consumers. For example, Emami-Naeini et al.~\cite{emami2020ask} found that consumers particularly wanted to know who their data might be shared with. Beyond label content, researchers have discussed four common consumer-facing label formats -- binary, graded, descriptive, and multi-layer -- each with distinct tradeoffs~\cite{caven2025usability, camp2021lessons, johnson2020impact, caven2022integrating}. Yet, in GenAI -- particularly around S\&P -- such empirical and design guidance on what information to prioritize and how best to present it remains scarce.

Current real-world examples of GenAI transparency -- though not necessarily S\&P-oriented -- include model cards~\cite{mitchell2019model}, system cards~\cite{meta_system_cards}, and datasheets~\cite{ibm_ai_factsheets_docs}, which are technical documentation that summarizes key properties of models and datasets for developers, auditors, and other experts. However, efforts to support \textit{consumer-facing} S\&P transparency in GenAI remain limited. Moreover, we lack empirical insights into whether and how GenAI users seek, interpret, and evaluate the S\&P information that is available today.

Policy has likewise offered limited support for consumer-facing S\&P disclosures in GenAI. In the U.S., there is no comprehensive federal requirement for standardized end-user disclosures, and emerging state laws largely emphasize organizational accountability (e.g., Colorado’s SB24-205~\cite{colorado_sb24_205_2024} focuses on risk management and documentation for ``high-risk'' AI in consequential decisions). The EU AI Act~\cite{eu_ai_act_2024} introduces user-facing transparency obligations (e.g., notice of AI interaction and of emotion recognition/biometric categorization), but does not specify a format to present these notices, leaving implementation uncertain. Therefore, it is crucial to inform policies with empirical and design insights that can guide GenAI companies’ S\&P transparency -- both in what information to disclose and how to disclose it.
\section{Methodology} \label{sec:methods}
\paragraphb{Recruitment}
We recruited participants from Prolific\footnote{\url{https://www.prolific.com/}} from August to October, 2025. The Prolific participants were directed to a Qualtrics\footnote{\url{https://www.qualtrics.com/}} screening survey. To mitigate social desirability~\cite{grimm2010social} and demand characteristics~\cite{mccambridge2012effects}, we advertised our study without mentioning its S\&P focus, broadly framing it as ``understanding users' perspectives and experiences with GenAI tools.'' To capture a comprehensive range of consumer-facing GenAI products, we defined ``GenAI tools'' in both the pre-screening survey and interviews to include ``standalone GenAI apps'' (e.g., ChatGPT) and ``GenAI features built into other platforms, apps, or smart devices'' (e.g., Alexa+).

\paragraphb{Eligibility and pre-screening} Participants eligible for the main study must (1) be at least 18 years old, (2) live in the United States, and (3) have used at least one GenAI tool. In the pre-screening survey, participants answered eligibility questions, provided consent to participate in the survey, indicated their willingness to participate in the 45-60 minute main study, which included an interview and design session on Zoom\footnote{\url{https://zoom.com/}}. Participants who were ineligible, did not provide full consent, or were not interested in the main study were redirected to Prolific and compensated at a flat rate of 0.15 USD. Participants who proceeded to the remaining survey reported their Prolific IDs, GenAI usage, and demographics. We required a minimum threshold of GenAI usage (i.e., having used at least one GenAI tool) to support richness of insights. All participants who completed the survey were compensated at an hourly rate of 14.26 USD. The median completion time was approximately 3 minutes. Complete recruitment and pre-screening materials are included in Appendix~\ref{sec:appendix_artifacts}.

\paragraphb{Pilot interviews} To assess interview clarity and timing, we conducted two pilot interviews with participants recruited through the first author’s personal connections. Both participants met the main study eligibility criteria. After each pilot, the research team debriefed on question clarity and any difficulties the participant encountered when answering. We then revised unclear questions, reordered prompts to better match a natural conversational flow, and prepared backup prompts for cases where participants struggled to respond.

\paragraphb{Main study procedure} To reduce no-shows, we messaged each participant on Prolific one day before the main study to confirm attendance and shared brief instructions for using Zoom Whiteboard\footnote{Getting started with Zoom Whiteboard: 
\url{https://support.zoom.com/hc/en/article?id=zm_kb&sysparm_article=KB0059671}.}. We reached data saturation~\cite{saunders2018saturation} at the seventeenth study, i.e., no additional insights emerged. Aligned with qualitative methodology guidance~\cite{francis2010adequate}, we conducted four additional studies and did not identify substantially new insights. In total, we conducted 21 interviews and 20 design sessions\footnote{One participant did not complete the design session due to time constraints.}. Each participant received 20 USD in compensation after completing the main study. Figure~\ref{fig:zoom} provides an overview of the main study procedure. Each session involved one primary interviewer and a second researcher taking notes. The full protocol is included in Appendix~\ref{sec:appendix_interview}. The main study consisted of the following sections:

\begin{itemize}[leftmargin=10pt, itemsep=0pt, topsep=0pt, parsep=0pt, partopsep=0pt]
\item \textbf{Section 0: Introduction \& GenAI definitions.} We asked participants to share their definitions of GenAI and restated our definition of ``GenAI tools'' to align our terminology.

\item \textbf{Section 1: Consumer-facing AI tool usage.} We asked participants about the GenAI tools they used, their purposes of use, any tools they had discontinued and their reasons for discontinuation, and the GenAI tool they used most.

\item \textbf{Section 2: Pre-adoption considerations.} To comprehensively capture factors shaping participants’ experiences across their entire use journey, we began by probing their pre-adoption considerations, including what criteria they used to choose a GenAI tool and what information they sought to support that decision.

\item \textbf{Section 3: Post-adoption considerations.} We then asked participants about the factors shaping their post-adoption use. To avoid priming S\&P concerns, we first invited participants to describe, in general terms, the perceived benefits and concerns. After they had shared all their initial thoughts, we followed up with questions about their specific S\&P considerations and any mitigation practices post-adoption.

\item \textbf{Section 4: Practices, challenges, \& expectations toward S\&P information seeking.} We asked participants about their S\&P information-seeking practices, including the sources they consulted, how they evaluated and felt about the information, and whether it led them to change their behavior or tool choices. We also asked about challenges they encountered and what additional information they wished they had to inform adoption and usage decisions.

\item \textbf{Section 5: Design sketching session.} After introducing the sketching tools, we asked participants to sketch transparency interfaces for the pre-adoption and post-adoption stages. We then asked participants to walk through their designs and explain their motivations and rationales.

\end{itemize}

\paragraphb{Qualitative data analysis} All interviews were audio-recorded and transcribed in English using Zoom’s transcription service. The first author proofread all transcripts prior to analysis to ensure accuracy and readability. Three researchers then conducted an iterative thematic analysis, consistent with prior studies~\cite{wei2024sok, liu2022your, nisenoff2023defining}. The first author served as the primary coder and developed two initial codebooks, one for the interview transcripts and one for participants’ design ideas. These codebooks were refined iteratively through weekly meetings with the full research team. Two secondary coders, one for the interview data and one for the design-session data, then applied the codebooks to 20\% of the dataset. We compared the primary and secondary coders' coding and computed inter-rater reliability (IRR). When agreement reached Cohen's $\kappa$ of 0.7, which is considered ``good''~\cite{blackman2000interval}, we proceeded using the primary coder's codes. When $\kappa<0.7$, the coders met to discuss and reconcile disagreements and to update the codebook. The secondary coders then applied the revised codebook to an additional 20\% of the data, repeating this process until Cohen’s $\kappa$ reached 0.7.

\paragraphb{Limitations} As is common in qualitative research, our sample size was relatively small. As such, our findings may not suggest established user needs at scale and may have limited generalizability in several ways. First, all participants were recruited via Prolific and were based in the U.S. Combined with our eligibility requirement of prior GenAI use, our sample likely over-represents relatively active GenAI users in the U.S. context. However, participants provided rich accounts of lived experiences, including concrete concerns and decision processes. This aligns with our exploratory goal of identifying where current S\&P transparency falls short in practice and what design improvements could better support users. Second, our study relied on self-reported accounts, which may be subject to social desirability bias~\cite{grimm2010social}. To mitigate this influence, we framed recruitment materials without mentioning ``security'' or ``privacy'' and recruited on a rolling basis to capture a diverse range of perspectives. Third, our intentionally broad definition of GenAI tools may limit the generalizability of our findings across specific tool types. Future work should further examine tool-specific transparency needs. Fourth, as users often develop S\&P considerations over time~\cite{li2023s}, we structured our interview questions and design tasks around pre- and post-adoption phases to capture a broader use journey rather than a single timepoint. However, this framing may have subtly shaped how participants organized their reflections across phases. To mitigate this potential bias, we emphasized throughout the interview and design session that participants could provide the same answers or propose the same design outcomes across both phases if they felt both applied. Finally, because GenAI S\&P transparency is an emerging and potentially unfamiliar design space, our use of lightweight prompts and interface-sketching tasks may have oriented participants toward interface-level solutions. Future work could use more open-ended design activities to surface additional forms of GenAI S\&P transparency.


\paragraphb{Ethics statement} This study was approved by our Institutional Review Board (IRB). All participants provided informed consent at each study stage. Participation was voluntary, and participants could skip any question, pause, or stop at any time without penalty. To minimize risk, we avoided requesting sensitive personal identifiers and informed participants that they could keep their camera off throughout the session. We audio-recorded sessions only with explicit consent. Before analysis, the first author proofread and de-identified transcripts and design sketches. Only the research team had access to the raw data. We report only de-identified quotes and descriptions and avoid including details that could reasonably re-identify participants.

\section{Results} \label{sec:results}
Our participants represented diverse demographics and exhibited varied levels of GenAI usage (Table~\ref{tab:participants}). We interviewed 21 U.S.-based GenAI users (P1--P21), ranging in age from 25 to 74, with most between 45 and 54 years old (11/21). 10 participants identified as Female, 10 identified as Male, and one identified as Non-binary / third gender. In terms of ethnicity, participants identified as White (13/21), Asian (5/21), Black or African American (4/21), and Middle Eastern or North African (1/21). Participants were generally frequent GenAI users: 18/21 used their most-used tool at least once a day. ChatGPT was the most commonly reported tool (13/21), followed by Gemini (5/21), Alexa+ (3/21), Copilot (2/21), DeepSeek (1/21), and Pi (1/21). Full demographics and GenAI usage information are included in Appendix~\ref{sec:appendix_artifacts}. In addition to reporting exact participant counts, we also use ``none'' (0\%), ``a few'' (1\%--24\%), ``some'' (25\%--44\%), ``about half'' (45\%--54\%), ``most'' (55\%--74\%), ``almost all'' (75\%--99\%), and ``all'' (100\%), where appropriate.

\begin{table*}[t]
  \centering
  \footnotesize
  \begin{tabular}{l p{4cm} l l p{3cm} p{4cm}}
    \toprule
    \textbf{ID} & \textbf{Most Often Used GenAI Tool(s)} & \textbf{Usage Frequency} & \textbf{Age} & \textbf{Gender} & \textbf{Ethnicity} \\
    \midrule
    P1 & Copilot & More than once a day & 45 - 54 & Female & White \\
    P2 & DeepSeek & More than once a day & 25 - 34 & Male & Middle Eastern or North African \\
    P3 & Alexa+, Gemini & More than once a day & 45 - 54 & Female & White \\
    P4 & ChatGPT & More than once a day & 45 - 54 & Male & Asian \\
    P5 & ChatGPT, Gemini & More than once a day & 45 - 54 & Female & White \\
    P6 & ChatGPT & More than once a day & 45 - 54 & Male & Asian \\
    P7 & Gemini & More than once a day & 35 - 44 & Female & Black or African American \\
    P8 & Alexa+ & More than once a day & 45 - 54 & Female & White \\
    P9 & ChatGPT & Once a week & 25 - 34 & Male & White \\
    P10 & ChatGPT & Once a week & 55 - 64 & Female & Asian \\
    P11 & Gemini & More than once a day & 65 - 74 & Male & White \\
    P12 & ChatGPT & More than once a day & 25 - 34 & Male & Asian \\
    P13 & Alexa+, ChatGPT & More than once a day & 45 - 54 & Male & White \\
    P14 & ChatGPT & Once a day & 45 - 54 & Male & White \\
    P15 & ChatGPT & Once a month & 35 - 44 & Female & Black or African American \\
    P16 & Pi, Copilot & Once a day & 35 - 44 & Female & Black \& White \\
    P17 & Gemini & More than once a day & 45 - 54 & Female & Asian \\
    P18 & ChatGPT & Once a day & 45 - 54 & Male & White \\
    P19 & ChatGPT & More than once a day & 45 - 54 & Male & White \\
    P20 & ChatGPT & Once a day & 25 - 34 & Female & Black \& White \\
    P21 & ChatGPT & Once a day & 25 - 34 & Non-binary / third gender & White \\
    \bottomrule
  \end{tabular}
  \caption{Summary of participant demographics and GenAI usage. Full table in Appendix~\ref{sec:appendix_artifacts}.}
  \label{tab:participants}
\end{table*}

\subsection{RQ1: Factors in GenAI Adoption \& Use} \label{sec:rq1}

\subsubsection{Factors in Adoption Decisions} \label{sec:rq1_1}
\paragraphb{Popularity, utility fit, and price as main drivers of initial adoption} The popularity of a GenAI tool was the primary factor shaping most participants' adoption decisions (14/21). For instance, P4 explained:
\begin{quote}
    ``[The primary factor that informed adoption] was just what was out there and being talked about, seeing that other people were getting useful results.''
\end{quote}

Notably, most participants who mentioned popularity (10/14) treated a tool's high popularity as a signal of stronger S\&P protections. Conversely, a few participants resisted adopting tools they perceived as less popular due to S\&P concerns (P4, P9). For example, P9 mentioned:

\begin{quote}
    ``A lot of these tools, like ChatGPT and whatnot, are so big that, if I end up having a problem, everybody else will have a problem, too. So they're big enough that I feel fairly safe using it. But if there's a name that seems suspicious to me [...] I just wouldn't even go for it. Reputable names are how I focus on being safe and secure, mostly.''
\end{quote}

Following popularity, context fit -- namely, how well a tool fits users' utility needs -- was another major driver of adoption (12/21). For example, P9 said:

\begin{quote}
    ``I'd want to see what benefit I'm going to get from the tool. How it would help me, what its capabilities or features are, and whether it aligns with a problem I'm having.''
\end{quote}

The third most commonly mentioned factor shaping users' adoption decisions was the price (7/21), with lower-cost or non-subscription tools perceived as more desirable. For example, P8 mentioned:

\begin{quote}
    ``How do I evaluate? Most of the time, I go on Google [...] I do a search and I look for prices, because I want to see what's cheaper [...] I don't want to pay anything monthly [...] That's a big factor, the price.''
\end{quote}

\paragraphb{S\&P certainty as a desired adoption factor, despite scarce information} When asked -- without any S\&P prompts -- what shaped their adoption decisions, some participants spontaneously mentioned S\&P as an important consideration at adoption (7/21). For instance, P5 mentioned:

\begin{quote}
    ``Initially, I was very hesitant to use any of them [GenAI tools] because I was concerned about privacy.''
\end{quote}

However, about half of participants reported that needed S\&P information about GenAI tools was not readily available or usable at the point of adoption (10/21). P21 said:

\begin{quote}
    ``If there was any AI that was very clear and upfront about things like, `we encrypt things,' or whatever, any company that was transparent about that stuff would be a better green flag [...] But the thing is, I don't see any of that with any of the AI currently [...] If they were really transparent and open about that, I'd be interested.''
\end{quote}



\subsubsection{Factors in Post-Adoption Experiences} \label{sec:rq1_2} 

\paragraphb{Output quality driving continued use of GenAI tools} Most participants noted that their continued use of GenAI tools was mainly driven by the perceived quality of outputs (12/21), including perceived accuracy/credibility (8/12) and lack of bias (4/12) in AI-generated information. Participants reported that tools that performed well on these metrics improved their efficiency, reduced cognitive load, or exposed them to more diverse, yet neutral, perspectives. For example, P2 said:

\begin{quote}
    ``The factors were, most importantly, performance, how quickly it works, and its accuracy as well [...] I try the same prompt on many platforms, and the one that gives me the most accurate and best results is the one I continue using.''
\end{quote}

Additionally, some participants cared about whether GenAI tools exhibited biased or overly opinionated behavior (4/12). For example, P16 mentioned:

\begin{quote}
    ``ChatGPT has bias. It has told me that the sources that it gets are verifiable news [...] but I've noticed it's been very biased towards, politically, a certain direction.''
\end{quote}

\paragraphb{S\&P uncertainty constraining AI use cases} About half of participants pointed out that limited visibility into GenAI tools' S\&P practices made it difficult to adopt concrete, effective risk mitigation strategies (10/21). As a result of S\&P uncertainty, a few participants chose to reduce or even discontinue their use of some GenAI tools (3/21), which in turn undermined the value of personalization. For example, P1 described how she significantly reduced her use of ChatGPT and stopped signing in due to a data breach~\cite{silberling2025publicChatGPTIndexed}:

\begin{quote}
    ``I've dropped off on ChatGPT a lot over the last week or two, when I found out that they had a leak and that people's chats were found on Google. That very much bothered me [...] Before, I would log in and let it track what I talked about so that I could get a more personalized experience. But if that means that I have to give up my privacy and what we talk about, it's not a good trade-off for me. Now I'm using it for very generic information, and I'm also not logging in.''
\end{quote}

Such constrained use became more pronounced when perceived S\&P stakes were higher. About half of participants reported heightened anxiety or discomfort when using GenAI tools in higher-stakes S\&P contexts (10/21), such as legal support (3/10), proprietary content creation (3/10), work-related tasks (3/10), and other sensitive uses. P10 worried that sensitive legal information, such as litigation strategy, could be exposed to opposing parties. When considering GenAI tools for legal support, P10 indicated that they would compare different tools' S\&P practices, if such information were available:

\begin{quote}
    ``For legal uses, which is a lot more sensitive, you would be comparing the privacy level of one against the other. In the case of a law firm, their approach to litigating, for example, you don't want that known to your opposing party -- it could cost a lot if somebody can get hold of the information and somehow use it in a nefarious way.''
\end{quote}

Additionally, some participants who worked in content creation raised concerns about content theft and expressed low trust in GenAI's training practices (3/10). P13 explained:

\begin{quote}
    ``If I was writing a movie script or a TV show that I wanted to keep proprietary, I don't trust that they're not going to use it in their models.''
\end{quote}

Some participants expressed uncertainty about whether workplace surveillance might monitor employees' GenAI tool use (3/10). For example, P6 said:

\begin{quote}
    ``Some AI tools are prohibited at our institute [...] So if I upload a file at work, and the file contains sensitive data [...] I'm not sure whether all these actions, like data input and file uploading, are monitored by the institute. That is the issue.''
\end{quote}


\subsection{RQ2: Current S\&P Information Seeking} \label{sec:rq2}

\subsubsection{Sources Used to Find S\&P Information} \label{sec:rq2_1}
\paragraphb{Skimming or summarizing information from official sources} Most participants reported consulting official documentation for S\&P information (15/21), either by skimming it themselves (13/15) or summarizing it using a GenAI tool (2/15). P5 described her information-seeking practice:

\begin{quote}
    ``I skimmed the terms and conditions on privacy information for ChatGPT and Gemini.''
\end{quote}

P20 described her approach of summarizing the information using a GenAI tool:

\begin{quote}
    ``Looking at the privacy statements they have, even though they make them very difficult and 50 or 60 pages long, I end up having to use AI to summarize it.''
\end{quote}

\paragraphb{Consulting trusted independent parties' opinions and lived experiences regarding GenAI tools' S\&P practices} Some participants reported consulting third-party sources to find information about GenAI tools' S\&P practices (6/21), including articles (3/6) and online communities (6/6). P2 described reading articles about Grok's S\&P practices, which led him to avoid using it:

\begin{quote}
    ``I read a lot of articles about Grok AI, for example, and how it works, and how it uses information. Basically, the information is used constantly. It's sold and given to many vendors [...] That's why I don't use it at all.''
\end{quote}

P4 reported consulting Reddit for other users' perspectives on GenAI tools' S\&P practices:

\begin{quote}
    ``I look at Reddit and see people who are in a similar position to me, using AI tools in similar ways, and what their thoughts were. It helps me get a general sense of what others think about these same topics [S\&P practices of GenAI tools].''
\end{quote}

Some of these participants compared multiple sources to evaluate S\&P information (2/6). P1 was interested in how S\&P practices worked in practice, comparing official documentation with other users' lived experiences shared online:

\begin{quote}
    ``I cross-reference that [official documentation] with what people have said online about the reality of how those things [S\&P practices] have played out. Because sometimes what those privacy statements are saying may not actually align with what people are experiencing.''
\end{quote}

Similarly, P13 relied on online posts to keep up with changes to official documentation:

\begin{quote}
    ``Google is changing its terms of service. Sometimes, on tech sites such as X and Reddit, people post and say the terms of service have changed and it's really bad. So I pay attention to things like that.''
\end{quote}

\subsubsection{Perceptions Toward Existing S\&P Transparency} \label{sec:rq2_2}

\paragraphb{Existing S\&P transparency perceived as ineffective} The perceived ineffectiveness was associated with two main reasons: (1) participants often did not know what they had agreed to due to lengthy, difficult language (6/21), and (2) changes to S\&P practices were not communicated in ways that participants could easily notice or access (3/21). P1 noted:

\begin{quote}
    ``I don't feel I've found the reality or the truth of the situation. When they have pages and pages of legal mumbo jumbo, it feels like a smokescreen to confuse people [...] It doesn't make me feel I have all the information I need to really feel comfortable.''
\end{quote}



A few participants expressed concerns about ineffective communication of changes to S\&P practices (3/21). P16 said:

\begin{quote}
    ``Sometimes there are pop-ups that say, `Hey, we've changed something.' But especially when you're logged in, if I don't open a completely new browser and I just pick back up in an old one and change the topic, it's not a fresh web page. So it doesn't automatically tell you if there's a new update. You have to exit and start a new session for that. So it's not transparent, and it doesn't notify you every time you visit the page.''
\end{quote}

\paragraphb{Existing S\&P transparency perceived as lacking credibility} Some participants were particularly concerned about the credibility of available S\&P information (8/21). P3 expressed doubt about official claims in the terms of service:

\begin{quote}
    ``There are a lot of claims that they try to make it as secure as possible, but I don't know if I believe it. Everything says, `We're good, and we protect your information,' but I still don't feel confident in their explanations about how they keep things safe for users.''
\end{quote}

P1 talked about how the scandal around ChatGPT users' chats appearing in public search results~\cite{silberling2025publicChatGPTIndexed} heightened her concerns about the credibility of official documentation across consumer-facing GenAI tools:

\begin{quote}
    ``Sure, they can say X, Y, and Z in their privacy policy, but ChatGPT also said X, Y, and Z in their privacy policy, and what the reality became, don't match up. And so you have to wonder in the back of your mind, is that going to happen with Gemini? Is that going to happen with Copilot? Is that going to happen with the ones that are embedded in my bank account?''
\end{quote}

A few participants were worried about the lack of credibility regarding the effectiveness or outcomes of S\&P controls provided by GenAI tools (3/21). For example, P14 was concerned about not knowing whether their S\&P choices were respected when using ChatGPT:

\begin{quote}
    ``If I direct it [ChatGPT] not to remember a session, I worry that I don't have any confirmation that it won't. If I say, `Disregard all session history,' I can't really tell whether it did or not.''
\end{quote}

P18 raised similar concerns about ChatGPT and Copilot:

\begin{quote}
    ``I've looked at the privacy configuration within ChatGPT and some in Copilot [...] But quite honestly, I don't have a lot of faith in those toggle switches, that they really offer the privacy they say they do.''
\end{quote}

\paragraphb{Perceived surface-level assurance at a glance} A few participants reported a sense of assurance from existing S\&P transparency, often without engaging with the details (2/21). For instance, P5 described feeling reassured by privacy policies even without understanding everything:

\begin{quote}
    ``It [privacy policy] made me feel a little bit better. I didn't really understand everything I was reading, and I quickly got bored with it, so I just thought, `Whatever, I'm just going to go with it. I think it'll be fine.'''
\end{quote}





\subsection{RQ3: S\&P Transparency for GenAI} \label{sec:rq3} 

\subsubsection{S\&P Transparency Content Wishlist} \label{sec:rq3_1}

\paragraphb{Who has access to user data} Some participants wanted to know whether their data could be accessed by others (9/21), including unknown third parties (8/9), models (2/9), or other users (2/9). For instance, P6 wanted to know whether their information was being shared with third parties:

\begin{quote}
    ``They should show whether the data will be shared, for example with other companies or institutions, and whether it will be shared with third parties.''
\end{quote}

A few participants wanted to know whether and how their data was transmitted, including to models beyond their awareness (2/9). As P1 mentioned:

\begin{quote}
    ``[It would be good to know] how many different AI systems are you feeding it [data] into? Are you taking it [data], say, if you're using Alexa+, and you [Alexa+] feed my information into your model, but then you also feed it into another model for comparison purposes? Now my information is in some other model that I'm not aware of, and I didn't agree to.''
\end{quote}



A few participants were also curious about whether their input data, if used for training, might later surface in outputs to other users (2/9). For example, P14 said:

\begin{quote}
    ``If I'm going through a legal problem and talking to it [GenAI tool] a lot about that, I'm interested in knowing whether it uses my sessions to learn from and then advise other people. I'm interested in how it takes someone's session and applies that to other people's sessions.''
\end{quote}

\paragraphb{Independent evaluations of S\&P practices} A few participants called for independent evaluations or audits of GenAI companies' S\&P practices that could provide users with more credible, digestible, and comparable insights (4/21). For example, P13 envisioned a third-party website that would translate companies' S\&P practices into usable consumer-facing summaries and comparisons:

\begin{table*}[!t]  
    \centering
    \footnotesize
    \def\arraystretch{0.9}
    \begin{tabular}{p{0.16\textwidth} p{0.34\textwidth} p{0.38\textwidth}} \toprule
       \textbf{Design Dimension} & \textbf{Description} & \textbf{Representative Example(s)}  \\ \midrule
        \textbf{Modality}: \textit{Interactive} & S\&P information formats that support on-demand user interaction. & AI-enabled privacy FAQs; interactive cards. \\
        \textbf{Modality}: \textit{Static} & S\&P information formats that are presented in fixed, non-interactive forms. & Standardized visual cues (labels). \\ \midrule
        \textbf{Granularity}: \textit{High} & Highly detailed information or control types. & Fine-grained permissions over data uses, recipients, and data types, with configurable time bounds (e.g., ``always,'' ``this session only''). \\
        \textbf{Granularity}: \textit{Low} & Less detailed information or control types. & Summaries of most needed data practices only. \\ \midrule
        \textbf{Explainability} & How clearly the interface communicates what an S\&P option means and what it enables. & A brief explanation of the toggled choice. \\
        \textbf{Reflectiveness} & How the interface supports engagement and reflection with S\&P information. & Pre-commit confirmation and editing. \\
        \midrule
        \textbf{Findability}: \textit{High} & S\&P information and controls are highlighted and easily located with textual, visual, or multimodal cues. & Clear navigation cues (e.g., links or walkthrough animation); standalone privacy icon/section. \\ \midrule
        \textbf{Timing}: \textit{Repeated} & S\&P information is repeatedly presented. & Per-launch warning screens. \\
        \textbf{Timing}: \textit{One-Time} & S\&P information is presented only once. & One-off disclaimers at sign-up. \\
        \textbf{Timing}: \textit{Persistent} & S\&P information is persistently presented. & Texts, toggles, badges, or banners that remain visible in the interface and surface critical S\&P information, such as the privacy settings currently enabled (e.g., training on/off). \\
        \bottomrule
    \end{tabular}
  \caption{Design dimensions, descriptions, and representative design ideas for S\&P transparency.}
    \label{tab:design_dimensions}
\end{table*}

\begin{quote}
    ``There should be an independent website that would audit all the companies. So if you had a question about what the terms of service really are, you could go to this website. You could even have AI distill it for you into one paragraph, or show it in a chart. Then you could compare your options as a consumer. Which one is the safest, which is the most secure, and which one isn't going to sell my data, or is going to sell my data? There isn't really a clearinghouse for that.''
\end{quote}

P4 specified which evaluation providers he would trust more, drawing an analogy to food certification that combines public oversight with reputable private accreditation:

\begin{quote}
    ``We need a combination of government involvement as well as accredited third parties [...] Think about certified organic food. The organization that certifies it has to be in good standing and have a reputation. I think both spaces, the government side and the private industry side, are beneficial. They both have weak spots, but together they complement each other. So it would have to be something like that, where maybe there was certification through the government and third-party certification as well. That would make me feel safe.''
\end{quote}

\paragraphb{Information on data storage, training, and inference} A few participants requested transparency into what data GenAI tools store, train on, or infer (5/21). For instance, P14 requested visibility into what types of data are stored, including metadata and inferred sensitive data:

\begin{quote}
    ``I would be interested in how much metadata it's storing, like time of use, usage patterns, phone number, address, political affiliation, and income level.''
\end{quote}

\subsubsection{Envisioned Design Practices for GenAI S\&P} \label{sec:rq3_2} We categorize participants' desired design practices for S\&P transparency into five dimensions: (1) \textit{modality}, (2) \textit{granularity}, (3) \textit{explainability and reflectiveness}, (4) \textit{findability}, and (5) \textit{timing}. Under each dimension, we present representative design ideas from participants. Table~\ref{tab:design_dimensions} summarizes the design dimensions and sub-dimensions, along with their descriptions and representative design ideas.

\paragraphb{Combining interactive and static disclosure modalities to lower understanding barriers} Some participants envisioned interactive, on-demand formats that would let them ask questions, clarify terms, and progressively drill down into details (8/20). Their sketches illustrated modalities such as AI-enabled privacy FAQs and expandable cards that reveal additional S\&P specifics and explanations on click. These interactive elements appeared most often in participants' pre-adoption designs, where they described wanting quick, situation-driven answers without having to parse lengthy documentation, while still retaining the option to explore deeper details. All participants proposed at least one design element in a static modality, where S\&P information is presented in a fixed, non-interactive format (20/20). Examples included visual cues, such as labels, that convey S\&P assurances at a glance. Participants described these elements as useful for quick pre-adoption screening and comparison, especially when they wanted a baseline understanding of data practices without digging into details.

\begin{figure*}
    \centering
    \includegraphics[width=1\linewidth]{img/prototype.pdf}
    \caption{We assemble participants' representative design ideas across the five dimensions into a single prototype spanning pre-adoption (left) and post-adoption (right) phases. In the pre-adoption phase, the prototype presents an independent audit badge in a static format (\textbf{A}), concise explanations (\textbf{B}) paired with granular permission controls (\textbf{C}), and an AI-enabled FAQ about data practices (\textbf{D}). After users make initial choices, the interface summarizes each option with brief explanations and allows users to edit or confirm their settings (\textbf{E}). In the post-adoption phase, users can quickly enable a privacy-preserving mode that remains visible in the interface (\textbf{F}) and view a persistent privacy reminder (\textbf{H}). Users can update privacy preferences via two clear visual entry points (\textbf{G}), and specify whether changes apply globally across sessions (\textbf{I}) or only to the current session (\textbf{J}). To illustrate traceability, we present examples in Figure~\ref{fig:original} showing how participants' original sketches correspond to features in the prototype.}
    \label{fig:design}
\end{figure*}

\paragraphb{Low granularity pre-adoption, high granularity post-adoption} In general, participants preferred low granularity pre-adoption (e.g., a brief summary of the salient S\&P information, as specified in Section~\ref{sec:rq3_1}), but wanted high granularity post-adoption (e.g., fine-grained consent options or controls). In particular, some participants indicated that existing GenAI tools fell short in providing sufficient permission controls, especially around purposes of data use, data recipients, and the types of data stored, trained on, or inferred (7/20). Importantly, some of these participants emphasized the need for configurable time bounds for these permissions (3/7). For example, they wanted global settings that apply to all sessions by default, while also allowing local, session-specific settings for particular permissions.

\paragraphb{Supporting S\&P decision-making through explainability and reflectiveness} Some participants suggested that toggling a consent option should trigger a brief explanation of what the choice enables, what data it affects, and any potential consequences (8/20). After reviewing the explanation, users should be able to confirm or revise the setting before it takes effect. Participants described these ideas as a way to mitigate long-standing issues of \textit{notice fatigue}~\cite{bohme2011security} and \textit{habituation}~\cite{anderson2014users, bravo2013your}, where users click through S\&P information without processing its content. By pairing pre-commit explanations with opportunities to confirm or revise choices, these designs create additional moments for users to engage with S\&P information and make more informed decisions.

\paragraphb{Improving the findability of S\&P information} About half of participants expressed a desire for higher visibility and easier access to information and controls (9/20). Examples included clear navigation cues that point users to where they can update settings during sign-up, as well as a standalone icon or dedicated section for privacy settings post-adoption.

\paragraphb{Presenting S\&P information at the right timing} Timing refers to when S\&P information is surfaced during use. Participants described three timing patterns -- \emph{repeated}, \emph{one-time}, and \emph{persistent}. Repeated disclosures include per-launch warning screens that caution users against sharing sensitive information. One-time disclosures include a one-off notice at sign-up and just-in-time notices triggered by a specific action. Persistent disclosures include texts, toggles, badges, or banners that remain visible in the interface and surface critical S\&P information, such as the privacy settings currently enabled (e.g., training on/off). For example, P13 specifically described a repeated, entry-point disclosure that appears at tool launch to provide a consistent reminder of data practices, analogous to cigarette warnings or car startup screens:

\begin{quote}
    ``When you open the [GenAI] app, there could be a warning that says, `Hey, this is what's going to happen to your data,' right when you open it, every time. That's like what we have with cigarettes. We have a label right on the pack [...] Now, in some cars, it gives you a warning screen right on the interface when you start the car up. Then it goes away, you hit OK, and you can drive. There might be something like that with an AI interface.''
\end{quote}

\section{Discussion} \label{sec:discussion}
\label{sec:discussion}
\paragraphb{Summary of findings} Our findings highlight an alarming status quo -- \textbf{S\&P transparency in consumer-facing GenAI tools is often missing or unusable in practice, even as high-stakes use becomes more common} (RQ1\&RQ2). In response to this challenge, participants described concrete expectations for \textbf{\textit{what} S\&P information should cover and \textit{how} it should be communicated} (RQ3). We organize participants' design ideas into five dimensions (Table~\ref{tab:design_dimensions}) and instantiate them in an interactive prototype (Figure~\ref{fig:design}) that visualizes some representative design ideas. For better readability of Figure~\ref{fig:design}, the interactive prototype can be accessed in Appendix~\ref{sec:appendix_artifacts}. Note that these design practices reflect participants' stated preferences rather than recommendations evaluated through observed behavior. The prototype is intended to surface an emerging design space of GenAI S\&P transparency by illustrating participant-informed design ideas. It is not meant to be comprehensive and does not necessarily reflect best practices. As an exploratory study, our goal is to offer a starting point for designing and evaluating GenAI S\&P transparency and to spur further investigation into best practices.

\paragraphb{Unique design challenges and opportunities for advancing GenAI S\&P transparency} Several unique S\&P transparency challenges emerged in the context of GenAI. First, our findings suggest structural design limitations that produce a mitigation-risk misalignment: privacy risks in GenAI tools can shift across use contexts, yet participants struggled to adjust their mitigation strategies accordingly for two main reasons. First, participants often used the same GenAI tool for both high-stakes and low-stakes purposes (Section~\ref{sec:rq1_2}). However, existing controls in GenAI tools are often coarse, global settings (e.g., a training opt-out/opt-in setting~\cite{openai_datacontrols_collection} that applies across sessions), making it difficult to manage risk as usage contexts shift. Second, participants perceived data practices as changing frequently without effective communication, leaving them unaware of what had changed and uncertain about current policies (Section~\ref{sec:rq2_1}). This uncertainty made it harder for participants to make informed S\&P decisions, such as what information to share or withhold. This challenge is particularly salient in GenAI because a single general-purpose tool is routinely used across contexts with very different risk levels (e.g., mental health support vs. information seeking~\cite{chatterji2025people}), whereas prior digital technologies often let users segment disclosure by setting or audience (e.g., social media platforms allow users to restrict a post's visibility to specific friend groups~\cite{liu2011analyzing}).

In response to these challenges, participants proposed design ideas across five dimensions (Section~\ref{sec:rq3_2}), several of which overlap with established dimensions in prior transparency design work, such as timing, channel, modality, and user control~\cite{schaub2015design}. Thus, our contribution is not to claim these dimensions as new, but to show how they may need to be reconfigured for GenAI use journeys. Because risk levels can shift quickly within the same tool, a single, fixed transparency approach is likely insufficient. Instead, for example, GenAI S\&P transparency may need to combine low-granularity summaries that support quick comparison and adoption decisions with high-granularity, session-specific explanations and controls that support higher-stakes use after adoption.


\begin{table*}[t!]
  \centering
  \setlength{\tabcolsep}{0pt} 
  \footnotesize

  \newcolumntype{C}{>{\centering\arraybackslash}p{1.8em}}

  \begin{tabular*}{\textwidth}{
    @{\extracolsep{\fill}}
    b{11em}       
    b{4em}        
    *{2}{C}       
    *{2}{C}       
    *{2}{C}       
    C             
    *{3}{C}       
    *{3}{C}       
    }

    &
    &
    \multicolumn{2}{c}{\textbf{Modality}} &
    \multicolumn{2}{c}{\textbf{Granularity}} &
    \multicolumn{2}{c}{\textbf{Expl. \& Refl.}} &
    \multicolumn{1}{c}{\textbf{\shortstack{Find-\\ability}}} &
    \multicolumn{3}{c}{\textbf{Timing}} &
    \multicolumn{3}{c}{\textbf{Content}} \\

    \cmidrule(lr){3-4} \cmidrule(lr){5-6} \cmidrule(lr){7-8} \cmidrule(lr){9-9} \cmidrule(lr){10-12} \cmidrule(lr){13-15}

    \textbf{Example} & \textbf{Audience} &
    \verttext{Interactive} & \verttext{Static} &
    \verttext{High Granularity} & \verttext{Low Granularity} &
    \verttext{Explainability} & \verttext{Reflectiveness} &
    \verttext{High Findability} &
    \verttext{Repeated} & \verttext{One-Time} & \verttext{Persistent} &
    \verttext{Indep. Evals} & \verttext{Data Types} & \verttext{Data Recipients} \\
    \midrule




    Google Model Cards~\cite{google_modelcards_website} &

    Experts &

    \leftrule{\quasi} & \yes &           

    \leftrule{\yes} & \quasi &        

    \leftrule{\yes} & \quasi &           

    \leftrule{\quasi} &               

    \leftrule{\quasi} & \quasi & \yes &     

    \leftrule{\quasi} & \yes & \yes \\ 



    Privy (Template-based)~\cite{lee2025privy} &

    Experts &

    \leftrule{\quasi} & \yes &           

    \leftrule{\yes} & \partmark &           

    \leftrule{\yes} & \yes &          

    \leftrule{\no} &                  

    \leftrule{\quasi} & \quasi & \yes &      

    \leftrule{\no} & \yes & \yes \\   



    Privy (LLM-based)~\cite{lee2025privy} &

    Experts &

    \leftrule{\yes} & \quasi &           

    \leftrule{\yes} & \partmark &           

    \leftrule{\yes} & \yes &          

    \leftrule{\no} &                  

    \leftrule{\quasi} & \quasi & \yes &      

    \leftrule{\no} & \yes & \yes \\   



    AI Nutrition Facts Label~\cite{twilio_ai_nutrition_facts_website} &

    Users &

    \leftrule{\quasi} & \yes &           

    \leftrule{\quasi} & \yes &           

    \leftrule{\yes} & \quasi &          

    \leftrule{\no} &                 

    \leftrule{\quasi} & \quasi & \yes &     

    \leftrule{\quasi} & \yes & \yes \\ 

    Manus Trust Center~\cite{manus_trust_center} &
    Users &
    \leftrule{\quasi} & \yes &           
    \leftrule{\yes} & \partmark &        
    \leftrule{\yes} & \quasi &           
    \leftrule{\quasi} &                 
    \leftrule{\quasi} & \quasi & \yes &     
    \leftrule{\yes} & \yes & \yes \\  

    \bottomrule
  \end{tabular*}

  \smallskip
  \begin{flushleft}\footnotesize
  \centering
    \textit{Evaluation}: \yes\ = example fulfills criterion; \partmark = example partially fulfills criterion; \quasi\ = example fails to fulfill criterion; \no\ = criterion not applicable.
  \end{flushleft}
 \vspace{-0.2in} \caption{Mapping a small subset of existing S\&P transparency approaches in GenAI to participant-informed design dimensions and information needs. Note: The table format is adapted from~\cite{stephenson2022sok}.}
  \label{fig:evaluation}
\end{table*}

Second, existing GenAI S\&P communication faces a significant ``credibility gap.'' While independent evaluations have been used to strengthen the credibility of self-attested S\&P practices in prior technologies (e.g., the FCC-led Cyber Trust Mark Program for IoT technologies~\cite{caven2025usability}), GenAI still largely lacks comparable, consumer-legible signals that reduce information asymmetry~\cite{morgner2018opinion} by clearly communicating what is being assessed, by whom, and what the results imply for users' data practices (Section~\ref{sec:rq3_1}). Moreover, policy efforts to advance S\&P transparency in GenAI remain limited and have yet to establish industry-wide standards for S\&P communication backed by trusted public authorities. Prior work further suggests that credibility-oriented S\&P disclosures (e.g., well-designed labels) can increase consumers' willingness to adopt or purchase privacy- and security-protective technologies~\cite{cranor2024internet, egelman2009timing, emami2023consumers, tsai2011effect}. However, in the GenAI context, labels may face additional credibility challenges because opaque data flows make it difficult for users to assess whether disclosed practices are accurate or complete. Building on this, future work should further examine (1) what content and forms of independent evaluation outputs (e.g., binary vs.\ graded, descriptive, or multi-layered labels~\cite{caven2025usability, chiang2025iot}) best align with GenAI users' needs, and (2) whether and how independent certifications shape users' adoption decisions and interactions in the GenAI context.





Despite these challenges, GenAI also introduces unique opportunities in the design space of S\&P transparency. Participants expressed a need for on-demand S\&P explanations delivered through interactive modalities (Section~\ref{sec:rq3_2}). Prior work on algorithmic transparency has similarly drawn on the UX principle of progressive disclosure~\cite{springer2020progressive, muralidhar2024effect, carroll1984training}, suggesting that providing explanations on an ``as needed'' basis can help users better understand a system and reduce initial errors. Our findings echo this preference in the S\&P context for GenAI (Section~\ref{sec:rq3_2}), particularly given participants' perceptions of ever-updating data practices in GenAI (Section~\ref{sec:rq2_1}). However, assessing its effectiveness in shaping users' S\&P understanding and behavior requires further empirical work.



\paragraphb{Evaluating existing GenAI S\&P transparency through our design dimensions} We compare a small set of existing S\&P transparency approaches in GenAI using our proposed design dimensions. Table~\ref{fig:evaluation} maps a subset of existing S\&P transparency approaches onto these dimensions\footnote{Two researchers independently audited the listed examples and resolved disagreements through discussion. We add \textit{content} as an additional table dimension to capture what information participants preferred to be included.}, and is intended to motivate future work that more systematically investigates and validates best practices for S\&P transparency in GenAI.

Existing S\&P transparency approaches in GenAI tools are largely not consumer-facing. Instead, they often target audiences with substantially higher technical expertise than typical end users. For instance, Google launched developer-facing model cards that help identify a GenAI model's strengths, limitations, and mitigations through text documentation~\cite{google_modelcards_website}.

Similarly, a recent study~\cite{lee2025privy} proposed \textit{Privy}, a practitioner-facing tool that guides AI product teams through privacy impact assessments (PIAs) in two versions -- static, structured vs. interactive, LLM-based. Their findings suggest that LLM-based PIAs identified more relevant and clear S\&P guidance compared to the static, structured version. Our findings also suggest participants' split preferences for interactive vs. static disclosure in different contexts.  Our participants wanted interactive disclosure to provide on-demand S\&P information and effective static disclosure to support quick pre-adoption screening and comparison. (Section~\ref{sec:rq3_2}).

Limited S\&P transparency approaches in GenAI have targeted general users, with the AI Nutrition Facts Label being one example~\cite{twilio_ai_nutrition_facts_website}. Similar to IoT label designs~\cite{emami2020ask}, the AI nutrition fact label presents a standardized summary of a GenAI model's data practices to help users quickly understand a model's data training, sharing, retention, and related practices. More recently, Manus~\cite{manus_trust_center} published consumer-facing information about its S\&P compliance, including independent audit reports, security controls, subprocessor disclosures, and static FAQs. Many of these emerging practices align with what our participants called for, further underscoring a growing need for consumer-facing transparency in GenAI.

In summary, our work introduces an initial design space for S\&P transparency in GenAI with \textit{user-informed} design practices, aiming to help designers, developers, and researchers identify and evaluate design choices when developing S\&P transparency for end users. To illustrate how this design space can be applied, Table~\ref{fig:evaluation} maps a small set of existing GenAI transparency approaches to participants' desired design dimensions and information needs. We emphasize that this mapping is not a systematic evaluation of the GenAI transparency landscape, and future work is needed to more comprehensively examine existing tools. However, this illustrative comparison suggests several emerging patterns across the selected examples, although these observations should not be interpreted as statistically reliable or representative of the broader GenAI transparency ecosystem. Across these examples, several approaches remain expert-facing, while consumer-facing disclosures are less developed. Interactive S\&P communication appears limited, with most examples relying on static formats. Timing is also largely persistent, with less support for repeated or one-time, just-in-time disclosures at decision points~\cite{egelman2009timing}. In addition, most examples convey S\&P information but provide fewer opportunities for users to reflect, confirm, or revise settings before higher-risk actions. Independent evaluations are also limited across the examples.



\paragraphb{Recommendations for future research: further validate, measure, and refine suggested design practices in real-world settings} Our work provides preliminary findings that surface exploratory design practices for promoting S\&P transparency in GenAI (Section~\ref{sec:rq3_2}). These practices should be further validated, measured, and refined through future empirical research. Future work should (1) evaluate the impact of GenAI S\&P transparency designs on users' perceptions and behaviors in real-world settings -- for example, how transparency granularity shapes users' understanding, decision-making, and disclosure practices as contexts and risk levels shift, and whether interactive, on-demand disclosure improves comprehension and reduces errors in practice; and (2) extend this work by translating participants' credibility concerns into policy-relevant evidence, examining what information and presentation formats for independent evaluation best support users' needs, which organizations (e.g., Consumer Reports\footnote{\url{https://www.consumerreports.org/}}, Underwriters Laboratories\footnote{\url{https://www.ul.com/}}) users trust more, and whether third-party certification measurably affects users' adoption and interaction behaviors in the GenAI context.

\paragraphb{Recommendations for future GenAI S\&P transparency design: balance flexibility and user burden through evidence-informed defaults} Participants called for more granular, flexible options that better match GenAI's shifting use contexts and risk levels (Section~\ref{sec:rq3_2}). At the same time, flexible controls can introduce trade-offs -- adding granularity can increase decision burden and may also create a false sense of safety that leads to greater disclosure over time~\cite{schaub2015design, brandimarte2013misplaced, keith2014privacy}. To balance these concerns, we recommend that designers make the default experiences secure, while offering additional granularity on demand. Additionally, prior work suggests that when data practices reasonably match users' expectations, they can demand less attention from users~\cite{schaub2015design}. Achieving this goal requires empirical research on users' mental models of GenAI data practices -- especially common misconceptions -- to inform defaults, explanations, and when (and how) to surface additional controls or information. Because average end users' S\&P knowledge may not be complete, future work should also incorporate technical and legal experts' perspectives when developing and evaluating GenAI S\&P transparency designs. In addition, these designs should be further evaluated through controlled and field experiments to understand their effects in practice.

\paragraphb{Recommendations for policies: help consumers gain industry-wide standardized insights} Participants suggested they would be more likely to trust independent evaluations that combine public oversight with accredited private auditors (Section~\ref{sec:rq3_1}), pointing to a role for policy in establishing industry-wide, consumer-legible S\&P transparency standards. Yet, legally enforceable transparency requirements for consumer-facing GenAI remain scarce within and beyond the U.S. In the U.S., pressure for truthful S\&P communication largely arrives through emerging state laws that focus on risk documentation for specific classes of systems (e.g., Colorado SB24-205’s duties and risk-management expectations for ``high-risk'' AI systems)~\cite{colorado_sb24_205_2024}. A key mismatch for our context is that these U.S.-based efforts tend to prioritize organizational accountability (e.g., internal governance processes and documentation)~\cite{nist_ai_rmf_2023, nist_genai_profile_2024}, while less often mandating consumer-legible disclosures about everyday data practices. Outside the U.S., transparency requirements also vary substantially in scope and intended audience -- for example, the EU AI Act introduces legally binding transparency obligations for certain AI systems and actors~\cite{eu_ai_act_2024}, while other efforts are largely non-binding, such as Singapore's Model AI Governance Framework for Gen AI~\cite{singapore_mgf_genai_2024} and the OECD's Hiroshima AI Process Reporting Framework~\cite{oecd_haip_reporting_2025}. Across these policy directions, ``transparency'' frequently emphasizes safety reporting and technical documentation rather than consumer-facing S\&P assurances.

Moving forward, policies could shift from primarily enforcing entity accountability to also mandating honest, consumer-facing S\&P transparency -- e.g., standardized, comparable disclosures and independent certification signals -- while guarding against ``transparency washing,'' such as vague or unverifiable claims, self-issued trust badges without auditable backing, or interface designs that nudge users into disclosure while obscuring critical terms~\cite{ftc_ai_comply_2024, ftc_dark_patterns_2022}.

\section{Conclusion} \label{sec:conclusion}
Our findings show that S\&P information rarely drove adoption, but S\&P uncertainty constrained high-stakes GenAI use. Participants found existing information sources ineffective or lacking credibility and wanted clearer disclosures. Their envisioned interfaces highlight five design dimensions, from which we derive recommendations for consumer-facing GenAI S\&P transparency.
\section*{Acknowledgments}
We thank the anonymous reviewers for their constructive feedback. We also thank our participants for their insightful input. This material is based upon work supported in part by a Google research gift award. Any opinions, findings, conclusions, or recommendations expressed in this material are those of the authors and do not necessarily reflect the views of the institutions.

{\footnotesize
\balance
\bibliographystyle{plain}
\bibliography{bibliography}}

@misc{silberling2025publicChatGPTIndexed,
  title        = {Your public ChatGPT queries are getting indexed by Google and other search engines},
  author = {Silberling, Amanda},
  howpublished = {\url{https://techcrunch.com/2025/07/31/your-public-chatgpt-queries-are-getting-indexed-by-google-and-other-search-engines/}},
  note         = {Accessed: 2026-02-05}
}

@inproceedings{harkous2016pribots,
  title={$\{$PriBots$\}$: Conversational privacy with chatbots},
  author={Harkous, Hamza and Fawaz, Kassem and Shin, Kang G and Aberer, Karl},
  booktitle={Twelfth Symposium on Usable Privacy and Security (SOUPS 2016)},
  year={2016}
}

@inproceedings{freiberger2026helping,
  title={Helping Johnny Make Sense of Privacy Policies with LLMs},
  author={Freiberger, Vincent and Fleig, Arthur and Buchmann, Erik},
  booktitle={Proceedings of the 2026 CHI Conference on Human Factors in Computing Systems},
  pages={1--21},
  year={2026}
}

@article{marchal2024generative,
  title={Generative AI misuse: A taxonomy of tactics and insights from real-world data},
  author={Marchal, Nahema and Xu, Rachel and Elasmar, Rasmi and Gabriel, Iason and Goldberg, Beth and Isaac, William},
  journal={arXiv preprint arXiv:2406.13843},
  year={2024}
}

@misc{manus_trust_center,
  author = {{Manus AI}},
  title = {Trust Center},
  year = {2025},
  url = {https://trust.manus.im/},
  note = {Accessed: 2025-02-02}
}

@misc{ftc_ai_comply_2024,
  author       = {{Federal Trade Commission}},
  title        = {FTC Announces Crackdown on Deceptive AI Claims and Schemes},
  howpublished = {Press Release},
  year         = {2024},
  month        = sep,
  url          = {https://www.ftc.gov/news-events/news/press-releases/2024/09/ftc-announces-crackdown-deceptive-ai-claims-schemes},
  urldate      = {2026-02-02}
}

@misc{colorado_sb24_205_2024,
  title        = {Colorado SB24-205: Consumer Protections for Artificial Intelligence},
  organization = {Colorado General Assembly},
  howpublished = {\url{https://leg.colorado.gov/bills/sb24-205}},
  note         = {Accessed: 2026-02-05}
}

@techreport{nist_ai_rmf_2023,
  author      = {{National Institute of Standards and Technology}},
  title       = {Artificial Intelligence Risk Management Framework (AI RMF 1.0)},
  institution = {NIST},
  year        = {2023},
  number      = {NIST AI 100-1},
  url         = {https://nvlpubs.nist.gov/nistpubs/ai/nist.ai.100-1.pdf},
  urldate     = {2026-02-02}
}

@inproceedings{chiang2025iot,
  title={IoT labels’ impact on security and privacy concerns},
  author={Chiang, Yi-Shyuan and Emami-Naeini, Pardis and Cobb, Camille},
  booktitle={2025 European Symposium on Usable Security (EuroUSEC)},
  pages={177--190},
  year={2025},
  organization={IEEE}
}

@techreport{nist_genai_profile_2024,
  author      = {{National Institute of Standards and Technology}},
  title       = {Artificial Intelligence Risk Management Framework: Generative Artificial Intelligence Profile},
  institution = {NIST},
  year        = {2024},
  number      = {NIST AI 600-1},
  url         = {https://nvlpubs.nist.gov/nistpubs/ai/NIST.AI.600-1.pdf},
  urldate     = {2026-02-02}
}

@misc{eu_ai_act_2024,
  author       = {{European Union}},
  title        = {Regulation (EU) 2024/1689 laying down harmonised rules on artificial intelligence (Artificial Intelligence Act)},
  howpublished = {EUR-Lex (Official Journal text)},
  year         = {2024},
  month        = jun,
  url          = {https://eur-lex.europa.eu/eli/reg/2024/1689/oj/eng},
  urldate      = {2026-02-02}
}

@techreport{singapore_mgf_genai_2024,
  author      = {{AI Verify Foundation}},
  title       = {Model AI Governance Framework for Generative AI},
  institution = {AI Verify Foundation (Singapore)},
  year        = {2024},
  month       = may,
  url         = {https://aiverifyfoundation.sg/wp-content/uploads/2024/05/Model-AI-Governance-Framework-for-Generative-AI-May-2024-1-1.pdf},
  urldate     = {2026-02-02}
}

@misc{oecd_haip_reporting_2025,
  title        = {OECD HAIP Reporting Framework},
  organization = {OECD},
  howpublished = {\url{https://transparency.oecd.ai/}},
  note         = {Accessed: 2026-02-05}
}

@misc{ftc_dark_patterns_2022,
  title        = {Bringing Dark Patterns to Light},
  organization = {Federal Trade Commission},
  howpublished = {\url{https://www.ftc.gov/reports/bringing-dark-patterns-light}},
  note         = {Accessed: 2026-02-05}
}

@article{liu2023prompt,
  title={Prompt injection attack against llm-integrated applications},
  author={Liu, Yi and Deng, Gelei and Li, Yuekang and Wang, Kailong and Wang, Zihao and Wang, Xiaofeng and Zhang, Tianwei and Liu, Yepang and Wang, Haoyu and Zheng, Yan and others},
  journal={arXiv preprint arXiv:2306.05499},
  year={2023}
}

@inproceedings{zhang2025understanding,
  title={Understanding Users' Privacy Perceptions Towards LLM's RAG-based Memory},
  author={Zhang, Shuning and Ma, Rongjun and Ma, Ying and Li, Shixuan and Xu, Yiqun and Yi, Xin and Li, Hewu},
  booktitle={Proceedings of the 2025 Workshop on Human-Centered AI Privacy and Security},
  pages={10--19},
  year={2025}
}

@inproceedings{wang2025mental,
  title={Mental Models of Generative AI Chatbot Ecosystems},
  author={Wang, Xingyi and Wang, Xiaozheng and Park, Sunyup and Yao, Yaxing},
  booktitle={Proceedings of the 30th International Conference on Intelligent User Interfaces},
  pages={1016--1031},
  year={2025}
}

@inproceedings{caven2022integrating,
  title={Integrating human intelligence to bypass information asymmetry in procurement decision-making},
  author={Caven, Peter J and Gopavaram, Shakthidhar Reddy and Camp, L Jean},
  booktitle={MILCOM 2022-2022 IEEE Military Communications Conference (MILCOM)},
  pages={687--692},
  year={2022},
  organization={IEEE}
}

@article{johnson2020impact,
  title={The impact of IoT security labelling on consumer product choice and willingness to pay},
  author={Johnson, Shane D and Blythe, John M and Manning, Matthew and Wong, Gabriel TW},
  journal={PloS one},
  volume={15},
  number={1},
  pages={e0227800},
  year={2020},
  publisher={Public Library of Science San Francisco, CA USA}
}

@inproceedings{camp2021lessons,
  title={Lessons for labeling from risk communication},
  author={Camp, Jean and Gopavaram, Shakthidhar and Dev, Jayati and Gumusel, Ece},
  booktitle={Workshop and Call for Papers on Cybersecurity Labeling Programs for Consumers: Internet of Things (IoT) Devices and Software},
  pages={1--3},
  year={2021},
  organization={NIST Washington DC}
}

@inproceedings{castelluccia2018enhancing,
  title={Enhancing transparency and consent in the IoT},
  author={Castelluccia, Claude and Cunche, Mathieu and Le M{\'e}tayer, Daniel and Morel, Victor},
  booktitle={2018 IEEE European Symposium on Security and Privacy Workshops (EuroS\&PW)},
  pages={116--119},
  year={2018},
  organization={IEEE}
}

@inproceedings{zimmeck2017automated,
  title={Automated Analysis of Privacy Requirements for Mobile Apps.},
  author={Zimmeck, Sebastian and Wang, Ziqi and Zou, Lieyong and Iyengar, Roger and Liu, Bin and Schaub, Florian and Wilson, Shomir and Sadeh, Norman M and Bellovin, Steven M and Reidenberg, Joel R},
  booktitle={NDSS},
  volume={2},
  pages={1--4},
  year={2017}
}

@article{sunyaev2015availability,
  title={Availability and quality of mobile health app privacy policies},
  author={Sunyaev, Ali and Dehling, Tobias and Taylor, Patrick L and Mandl, Kenneth D},
  journal={Journal of the American Medical Informatics Association},
  volume={22},
  number={e1},
  pages={e28--e33},
  year={2015},
  publisher={Oxford University Press}
}

@misc{meta_system_cards,
  title        = {Meta System Cards},
  organization = {Meta},
  howpublished = {\url{https://ai.meta.com/tools/system-cards/}},
  note         = {Accessed: 2026-02-05}
}

@misc{ibm_ai_factsheets_docs,
  title        = {IBM AI Factsheets},
  organization = {IBM},
  howpublished = {\url{https://www.ibm.com/docs/en/software-hub/5.1.x?topic=services-ai-factsheets}},
  note         = {Accessed: 2026-02-05}
}

@article{schwartz2004paradox,
  title={The paradox of choice: Why more is less HarperCollins Publishers},
  author={Schwartz, B},
  journal={New York, NY},
  year={2004}
}

@inproceedings{nisenoff2023defining,
  title={Defining" Broken": User Experiences and Remediation Tactics When $\{$Ad-Blocking$\}$ or $\{$Tracking-Protection$\}$ Tools Break a $\{$Website’s$\}$ User Experience},
  author={Nisenoff, Alexandra and Borem, Arthur and Pickering, Madison and Nakanishi, Grant and Thumpasery, Maya and Ur, Blase},
  booktitle={32nd USENIX Security Symposium (USENIX Security 23)},
  pages={3619--3636},
  year={2023}
}

@inproceedings{liu2022your,
  title={How Are Your Zombie Accounts? Understanding Users' Practices and Expectations on Mobile App Account Deletion},
  author={Liu, Yijing and Jia, Yan and Tan, Qingyin and Liu, Zheli and Xing, Luyi},
  booktitle={31st USENIX Security Symposium (USENIX Security 22)},
  pages={863--880},
  year={2022}
}

@article{blackman2000interval,
  title={Interval estimation for Cohen's kappa as a measure of agreement},
  author={Blackman, Nicole J-M and Koval, John J},
  journal={Statistics in medicine},
  volume={19},
  number={5},
  pages={723--741},
  year={2000},
  publisher={Wiley Online Library}
}

@inproceedings{wei2024sok,
  title={$\{$SoK$\}$(or $\{$SoLK?)$\}$: On the Quantitative Study of Sociodemographic Factors and Computer Security Behaviors},
  author={Wei, Miranda and Mink, Jaron and Eiger, Yael and Kohno, Tadayoshi and Redmiles, Elissa M and Roesner, Franziska},
  booktitle={33rd USENIX Security Symposium (USENIX Security 24)},
  pages={7011--7030},
  year={2024}
}

@inbook{grimm2010social,
author = {Grimm, Pamela},
publisher = {John Wiley \& Sons, Ltd},
isbn = {9781444316568},
title = {Social Desirability Bias},
booktitle = {Wiley International Encyclopedia of Marketing},
chapter = {},
pages = {},
doi = {https://doi.org/10.1002/9781444316568.wiem02057},
url = {https://onlinelibrary.wiley.com/doi/abs/10.1002/9781444316568.wiem02057},
eprint = {https://onlinelibrary.wiley.com/doi/pdf/10.1002/9781444316568.wiem02057},
year = {2010}
}

@article{mccambridge2012effects,
  title={The effects of demand characteristics on research participant behaviours in non-laboratory settings: a systematic review},
  author={McCambridge, Jim and De Bruin, Marijn and Witton, John},
  journal={PloS one},
  volume={7},
  number={6},
  pages={e39116},
  year={2012},
  publisher={Public Library of Science San Francisco, USA}
}

@article{murmann2021privacy_notifications,
  author  = {Patrick Murmann and Farzaneh Karegar},
  title   = {From Design Requirements to Effective Privacy Notifications: Empowering Users of Online Services to Make Informed Decisions},
  journal = {International Journal of Human-Computer Interaction},
  volume  = {37},
  number  = {19},
  pages   = {1823--1848},
  year    = {2021},
  doi     = {10.1080/10447318.2021.1913859},
  url     = {https://doi.org/10.1080/10447318.2021.1913859}
}

@inproceedings{kelley2009nutrition,
  title={A" nutrition label" for privacy},
  author={Kelley, Patrick Gage and Bresee, Joanna and Cranor, Lorrie Faith and Reeder, Robert W},
  booktitle={Proceedings of the 5th Symposium on Usable Privacy and Security},
  pages={1--12},
  year={2009}
}

@article{wogalter2002based,
  title={based guidelines for warning design and evaluation},
  author={Wogalter, Michael S and Conzola, Vincent C and Smith-Jackson, Tonya L},
  journal={Applied ergonomics},
  volume={33},
  number={3},
  pages={219--230},
  year={2002},
  publisher={Elsevier}
}

@inproceedings{good2007noticing,
  title={Noticing notice: a large-scale experiment on the timing of software license agreements},
  author={Good, Nathaniel S and Grossklags, Jens and Mulligan, Deirdre K and Konstan, Joseph A},
  booktitle={Proceedings of the SIGCHI conference on Human factors in computing systems},
  pages={607--616},
  year={2007}
}

@inproceedings{ali2025understanding,
  title={Understanding Users' Security and Privacy Concerns and Attitudes Towards Conversational AI Platforms},
  author={Ali, Mutahar and Arunasalam, Arjun and Farrukh, Habiba},
  booktitle={2025 IEEE Symposium on Security and Privacy (SP)},
  pages={298--316},
  year={2025},
  organization={IEEE}
}

@inproceedings{malki2025hoovered,
  title={``Hoovered up as a data point'': Exploring Privacy Behaviours, Awareness, and Concerns Among UK Users of LLM-based Conversational Agents},
  author={Malki, Lisa Mekioussa and others},
  booktitle={Proceedings on Privacy Enhancing Technologies},
  year={2025},
  organization={ACM}
}

@inproceedings{stephenson2022sok,
  title={Sok: Authentication in augmented and virtual reality},
  author={Stephenson, Sophie and Pal, Bijeeta and Fan, Stephen and Fernandes, Earlence and Zhao, Yuhang and Chatterjee, Rahul},
  booktitle={2022 IEEE symposium on security and privacy (SP)},
  pages={267--284},
  year={2022},
  organization={IEEE}
}

@article{tsai2011effect,
  title={The effect of online privacy information on purchasing behavior: An experimental study},
  author={Tsai, Janice Y and Egelman, Serge and Cranor, Lorrie and Acquisti, Alessandro},
  journal={Information systems research},
  volume={22},
  number={2},
  pages={254--268},
  year={2011},
  publisher={Informs}
}

@inproceedings{emami2023consumers,
  title={Are Consumers Willing to Pay for Security and Privacy of $\{$IoT$\}$ Devices?},
  author={Emami-Naeini, Pardis and Dheenadhayalan, Janarth and Agarwal, Yuvraj and Cranor, Lorrie Faith},
  booktitle={32nd USENIX Security Symposium (USENIX Security 23)},
  pages={1505--1522},
  year={2023}
}

@inproceedings{egelman2009timing,
  title={Timing is everything? The effects of timing and placement of online privacy indicators},
  author={Egelman, Serge and Tsai, Janice and Cranor, Lorrie Faith and Acquisti, Alessandro},
  booktitle={Proceedings of the SIGCHI Conference on Human Factors in Computing Systems},
  pages={319--328},
  year={2009}
}

@article{cranor2024internet,
  title={Internet of things security and privacy labels should empower consumers},
  author={Cranor, Lorrie Faith and Agarwal, Yuvraj and Emami-Naeini, Pardis},
  journal={Communications of the ACM},
  volume={67},
  number={3},
  pages={29--31},
  year={2024},
  publisher={ACM New York, NY, USA}
}

@inproceedings{morgner2018opinion,
  title={Opinion: Security lifetime labels-Overcoming information asymmetry in security of IoT consumer products},
  author={Morgner, Philipp and Freiling, Felix and Benenson, Zinaida},
  booktitle={Proceedings of the 11th ACM Conference on Security \& Privacy in Wireless and Mobile Networks},
  pages={208--211},
  year={2018}
}

@inproceedings{liu2011analyzing,
  title={Analyzing facebook privacy settings: user expectations vs. reality},
  author={Liu, Yabing and Gummadi, Krishna P and Krishnamurthy, Balachander and Mislove, Alan},
  booktitle={Proceedings of the 2011 ACM SIGCOMM conference on Internet measurement conference},
  pages={61--70},
  year={2011}
}

@techreport{chatterji2025people,
  title={How people use chatgpt},
  author={Chatterji, Aaron and Cunningham, Thomas and Deming, David J and Hitzig, Zoe and Ong, Christopher and Shan, Carl Yan and Wadman, Kevin},
  year={2025},
  institution={National Bureau of Economic Research}
}

@misc{openai_datacontrols_collection,
  title        = {ChatGPT Data Controls},
  organization = {OpenAI},
  howpublished = {\url{https://help.openai.com/en/collections/8471418-data-controls}},
  note         = {Accessed: 2026-02-05}
}

@String{Computing = "Computing" }

@String{Computer = "{IEEE} Computer" }

@String{Springer = "Springer-Verlag" }

@article{saunders2018saturation,
  title={Saturation in qualitative research: exploring its conceptualization and operationalization},
  author={Saunders, Benjamin and Sim, Julius and Kingstone, Tom and Baker, Shula and Waterfield, Jackie and Bartlam, Bernadette and Burroughs, Heather and Jinks, Clare},
  journal={Quality \& quantity},
  volume={52},
  number={4},
  pages={1893--1907},
  year={2018},
  publisher={Springer}
}

@inproceedings{bohme2011security,
  title={The security cost of cheap user interaction},
  author={B{\"o}hme, Rainer and Grossklags, Jens},
  booktitle={Proceedings of the 2011 New Security Paradigms Workshop},
  pages={67--82},
  year={2011}
}

@article{anderson2014users,
  title={Users aren’t (necessarily) lazy: Using neurois to explain habituation to security warnings},
  author={Anderson, Bonnie and Vance, Anthony and Kirwan, Brock and Eargle, David and Howard, Seth},
  year={2014}
}

@inproceedings{bravo2013your,
  title={Your attention please: Designing security-decision UIs to make genuine risks harder to ignore},
  author={Bravo-Lillo, Cristian and Komanduri, Saranga and Cranor, Lorrie Faith and Reeder, Robert W and Sleeper, Manya and Downs, Julie and Schechter, Stuart},
  booktitle={Proceedings of the Ninth Symposium on Usable Privacy and Security},
  pages={1--12},
  year={2013}
}

@article{balebako2014your,
  title={Is your inseam a biometric? a case study on the role of usability studies in developing public policy},
  author={Balebako, Rebecca and Shay, Richard and Cranor, Lorrie Faith},
  journal={Proc. USEC},
  volume={14},
  number={10.14722},
  year={2014}
}

@article{cranor2012necessary,
  title={Necessary but not sufficient: Standardized mechanisms for privacy notice and choice},
  author={Cranor, Lorrie Faith},
  journal={J. on Telecomm. \& High Tech. L.},
  volume={10},
  pages={273},
  year={2012},
  publisher={HeinOnline}
}

@inproceedings{jensen2004privacy,
  title={Privacy policies as decision-making tools: an evaluation of online privacy notices},
  author={Jensen, Carlos and Potts, Colin},
  booktitle={Proceedings of the SIGCHI conference on Human Factors in Computing Systems},
  pages={471--478},
  year={2004}
}

@article{fu2014field,
  title={A field study of run-time location access disclosures on android smartphones},
  author={Fu, Huiqing and Yang, Yulong and Shingte, Nileema and Lindqvist, Janne and Gruteser, Marco},
  journal={Proc. USEC},
  volume={14},
  number={10},
  year={2014}
}

@article{cranor2006user,
  title={User interfaces for privacy agents},
  author={Cranor, Lorrie Faith and Guduru, Praveen and Arjula, Manjula},
  journal={ACM Transactions on Computer-Human Interaction (TOCHI)},
  volume={13},
  number={2},
  pages={135--178},
  year={2006},
  publisher={ACM New York, NY, USA}
}

@inproceedings{kelley2010standardizing,
  title={Standardizing privacy notices: an online study of the nutrition label approach},
  author={Kelley, Patrick Gage and Cesca, Lucian and Bresee, Joanna and Cranor, Lorrie Faith},
  booktitle={Proceedings of the SIGCHI Conference on Human factors in Computing Systems},
  pages={1573--1582},
  year={2010}
}

@inproceedings{kelley2013privacy,
  title={Privacy as part of the app decision-making process},
  author={Kelley, Patrick Gage and Cranor, Lorrie Faith and Sadeh, Norman},
  booktitle={Proceedings of the SIGCHI conference on human factors in computing systems},
  pages={3393--3402},
  year={2013}
}

@article{brandimarte2013misplaced,
  title={Misplaced confidences: Privacy and the control paradox},
  author={Brandimarte, Laura and Acquisti, Alessandro and Loewenstein, George},
  journal={Social psychological and personality science},
  volume={4},
  number={3},
  pages={340--347},
  year={2013},
  publisher={Sage Publications Sage CA: Los Angeles, CA}
}

@inproceedings{keith2014privacy,
  title={Privacy fatigue: The effect of privacy control complexity on consumer electronic information disclosure},
  author={Keith, Mark J and Maynes, Courtenay and Lowry, Paul Benjamin and Babb, Jeffry},
  booktitle={International Conference on Information Systems (ICIS 2014), Auckland, New Zealand, December},
  pages={14--17},
  year={2014}
}

@misc{google_modelcards_website,
  title        = {Google Model Cards},
  organization = {Google},
  howpublished = {\url{https://modelcards.withgoogle.com/}},
  note         = {Accessed: 2026-01-21}
}

@misc{twilio_ai_nutrition_facts_website,
  title        = {AI Nutrition Facts},
  organization = {Twilio},
  howpublished = {\url{https://nutrition-facts.ai/}},
  note         = {Accessed: 2026-01-21}
}

@misc{lee2025privy,
  title         = {Privy: Envisioning and Mitigating Privacy Risks for Consumer-facing AI Product Concepts},
  author        = {Lee, Hao-Ping and Yang, Yu-Ju and Bilik, Matthew and Krsek, Isadora and Serban von Davier, Thomas and Monteiro, Kyzyl and Lin, Jason and Agarwal, Shivani and Forlizzi, Jodi and Das, Sauvik},
  year          = {2025},
  month         = sep,
  eprint        = {2509.23525},
  archivePrefix = {arXiv},
  primaryClass  = {cs.HC},
  doi           = {10.48550/arXiv.2509.23525},
  howpublished  = {\url{https://arxiv.org/abs/2509.23525}},
  note          = {Accessed: 2026-01-21}
}

@inproceedings{muralidhar2024effect,
  title={The Effect of Progressive Disclosure in the Transparency of Large Language Models},
  author={Muralidhar, Deepa and Belloum, Rafik and de Oliveira, Kathia Mar{\c{c}}al and Ashok, Ashwin and Mohammad, Pardaz Banu},
  booktitle={International Conference on Computer-Human Interaction Research and Applications},
  pages={269--288},
  year={2024},
  organization={Springer}
}

@article{carroll1984training,
  title={Training wheels in a user interface},
  author={Carroll, John M and Carrithers, Caroline},
  journal={Communications of the ACM},
  volume={27},
  number={8},
  pages={800--806},
  year={1984},
  publisher={ACM New York, NY, USA}
}

@article{springer2020progressive,
  title={Progressive disclosure: When, why, and how do users want algorithmic transparency information?},
  author={Springer, Aaron and Whittaker, Steve},
  journal={ACM Transactions on Interactive Intelligent Systems (TiiS)},
  volume={10},
  number={4},
  pages={1--32},
  year={2020},
  publisher={ACM New York, NY, USA}
}

@inproceedings{caven2025usability,
  title={Usability, Efficacy, and Acceptability of the US Cyber Trust Mark},
  author={Caven, Peter and Gurjar, Ambarish and Zhang, Zitao and Ma, Xinyao and Camp, LJean},
  booktitle={Proceedings of the 2025 CHI Conference on Human Factors in Computing Systems},
  pages={1--35},
  year={2025}
}

@inproceedings{langheinrich2002privacy,
  title={A privacy awareness system for ubiquitous computing environments},
  author={Langheinrich, Marc},
  booktitle={international conference on Ubiquitous Computing},
  pages={237--245},
  year={2002},
  organization={Springer}
}

@article{francis2010adequate,
  title={What is an adequate sample size? Operationalising data saturation for theory-based interview studies},
  author={Francis, Jill J and Johnston, Marie and Robertson, Clare and Glidewell, Liz and Entwistle, Vikki and Eccles, Martin P and Grimshaw, Jeremy M},
  journal={Psychology and health},
  volume={25},
  number={10},
  pages={1229--1245},
  year={2010},
  publisher={Taylor \& Francis}
}

@inproceedings{schneiders2025objection,
  title={Objection Overruled! Lay People can Distinguish Large Language Models from Lawyers, but still Favour Advice from an LLM},
  author={Schneiders, Eike and Seabrooke, Tina and Krook, Joshua and Hyde, Richard and Leesakul, Natalie and Clos, Jeremie and Fischer, Joel E},
  booktitle={Proceedings of the 2025 CHI Conference on Human Factors in Computing Systems},
  pages={1--14},
  year={2025}
}

@inproceedings{ma2025privacy,
  title={Privacy Perceptions of Custom GPTs by Users and Creators},
  author={Ma, Rongjun and Maidhof, Caterina and Carrillo, Juan Carlos and Lindqvist, Janne and Such, Jose},
  booktitle={Proceedings of the 2025 CHI Conference on Human Factors in Computing Systems},
  pages={1--18},
  year={2025}
}

@inproceedings{zhang2024s,
  title={“It's a Fair Game”, or Is It? Examining How Users Navigate Disclosure Risks and Benefits When Using LLM-Based Conversational Agents},
  author={Zhang, Zhiping and Jia, Michelle and Lee, Hao-Ping and Yao, Bingsheng and Das, Sauvik and Lerner, Ada and Wang, Dakuo and Li, Tianshi},
  booktitle={Proceedings of the 2024 CHI Conference on Human Factors in Computing Systems},
  pages={1--26},
  year={2024}
}

@article{tao2023opening,
  title={Opening a Pandora's box: things you should know in the era of custom GPTs},
  author={Tao, Guanhong and Cheng, Siyuan and Zhang, Zhuo and Zhu, Junmin and Shen, Guangyu and Zhang, Xiangyu},
  journal={arXiv preprint arXiv:2401.00905},
  year={2023}
}

@inproceedings{bird2023typology,
  title={Typology of risks of generative text-to-image models},
  author={Bird, Charlotte and Ungless, Eddie and Kasirzadeh, Atoosa},
  booktitle={Proceedings of the 2023 AAAI/ACM Conference on AI, Ethics, and Society},
  pages={396--410},
  year={2023}
}

@inproceedings{weidinger2022taxonomy,
  title={Taxonomy of risks posed by language models},
  author={Weidinger, Laura and Uesato, Jonathan and Rauh, Maribeth and Griffin, Conor and Huang, Po-Sen and Mellor, John and Glaese, Amelia and Cheng, Myra and Balle, Borja and Kasirzadeh, Atoosa and others},
  booktitle={Proceedings of the 2022 ACM conference on fairness, accountability, and transparency},
  pages={214--229},
  year={2022}
}

@article{kshetri2023cybercrime,
  title={Cybercrime and privacy threats of large language models},
  author={Kshetri, Nir},
  journal={IT Professional},
  volume={25},
  number={3},
  pages={9--13},
  year={2023},
  publisher={IEEE}
}

@inproceedings{carlini2022quantifying,
  title={Quantifying memorization across neural language models},
  author={Carlini, Nicholas and Ippolito, Daphne and Jagielski, Matthew and Lee, Katherine and Tramer, Florian and Zhang, Chiyuan},
  booktitle={The Eleventh International Conference on Learning Representations},
  year={2022}
}

@inproceedings{carlini2021extracting,
  title={Extracting training data from large language models},
  author={Carlini, Nicholas and Tramer, Florian and Wallace, Eric and Jagielski, Matthew and Herbert-Voss, Ariel and Lee, Katherine and Roberts, Adam and Brown, Tom and Song, Dawn and Erlingsson, Ulfar and others},
  booktitle={30th USENIX security symposium (USENIX Security 21)},
  pages={2633--2650},
  year={2021}
}

@article{zhang2023counterfactual,
  title={Counterfactual memorization in neural language models},
  author={Zhang, Chiyuan and Ippolito, Daphne and Lee, Katherine and Jagielski, Matthew and Tram{\`e}r, Florian and Carlini, Nicholas},
  journal={Advances in Neural Information Processing Systems},
  volume={36},
  pages={39321--39362},
  year={2023}
}

@article{van2019chatbot,
  title={Chatbot advertising effectiveness: When does the message get through?},
  author={Van den Broeck, Evert and Zarouali, Brahim and Poels, Karolien},
  journal={Computers in Human Behavior},
  volume={98},
  pages={150--157},
  year={2019},
  publisher={Elsevier}
}

@inproceedings{lee2024deepfakes,
  title={Deepfakes, phrenology, surveillance, and more! a taxonomy of ai privacy risks},
  author={Lee, Hao-Ping and Yang, Yu-Ju and Von Davier, Thomas Serban and Forlizzi, Jodi and Das, Sauvik},
  booktitle={Proceedings of the 2024 CHI Conference on Human Factors in Computing Systems},
  pages={1--19},
  year={2024}
}

@article{wang2024machine,
  title={Machine unlearning: A comprehensive survey},
  author={Wang, Weiqi and Tian, Zhiyi and Zhang, Chenhan and Yu, Shui},
  journal={arXiv preprint arXiv:2405.07406},
  year={2024}
}

@article{xu2024machine,
  title={Machine unlearning: Solutions and challenges},
  author={Xu, Jie and Wu, Zihan and Wang, Cong and Jia, Xiaohua},
  journal={IEEE Transactions on Emerging Topics in Computational Intelligence},
  year={2024},
  publisher={IEEE}
}

@article{zhang2023review,
  title={A review on machine unlearning},
  author={Zhang, Haibo and Nakamura, Toru and Isohara, Takamasa and Sakurai, Kouichi},
  journal={SN Computer Science},
  volume={4},
  number={4},
  pages={337},
  year={2023},
  publisher={Springer}
}

@inproceedings{schaub2015design,
  title={A design space for effective privacy notices},
  author={Schaub, Florian and Balebako, Rebecca and Durity, Adam L and Cranor, Lorrie Faith},
  booktitle={Eleventh symposium on usable privacy and security (SOUPS 2015)},
  pages={1--17},
  year={2015}
}

@inproceedings{emami2020ask,
  title={Ask the experts: What should be on an IoT privacy and security label?},
  author={Emami-Naeini, Pardis and Agarwal, Yuvraj and Cranor, Lorrie Faith and Hibshi, Hanan},
  booktitle={2020 IEEE Symposium on Security and Privacy (SP)},
  pages={447--464},
  year={2020},
  organization={IEEE}
}

@inproceedings{emami2021privacy,
  title={Which privacy and security attributes most impact consumers’ risk perception and willingness to purchase IoT devices?},
  author={Emami-Naeini, Pardis and Dheenadhayalan, Janarth and Agarwal, Yuvraj and Cranor, Lorrie Faith},
  booktitle={2021 IEEE Symposium on Security and Privacy (SP)},
  pages={519--536},
  year={2021},
  organization={IEEE}
}

@article{emami2021informative,
  title={An informative security and privacy “nutrition” label for internet of things devices},
  author={Emami-Naeini, Pardis and Dheenadhayalan, Janarth and Agarwal, Yuvraj and Cranor, Lorrie Faith},
  journal={IEEE Security \& Privacy},
  volume={20},
  number={2},
  pages={31--39},
  year={2021},
  publisher={IEEE}
}

@inproceedings{mitchell2019model,
  title={Model cards for model reporting},
  author={Mitchell, Margaret and Wu, Simone and Zaldivar, Andrew and Barnes, Parker and Vasserman, Lucy and Hutchinson, Ben and Spitzer, Elena and Raji, Inioluwa Deborah and Gebru, Timnit},
  booktitle={Proceedings of the conference on fairness, accountability, and transparency},
  pages={220--229},
  year={2019}
}

@article{mcdonald2008cost,
  title={The cost of reading privacy policies},
  author={McDonald, Aleecia M and Cranor, Lorrie Faith},
  journal={Isjlp},
  volume={4},
  pages={543},
  year={2008},
  publisher={HeinOnline}
}

@inproceedings{kwesi2025exploring,
  title={Exploring user security and privacy attitudes and concerns toward the use of $\{$General-Purpose$\}$$\{$LLM$\}$ chatbots for mental health},
  author={Kwesi, Jabari and Cao, Jiaxun and Manchanda, Riya and Emami-Naeini, Pardis},
  booktitle={34th USENIX Security Symposium (USENIX Security 25)},
  pages={6007--6024},
  year={2025}
}

@inproceedings{li2023s,
  title={“It’s up to the Consumer to be Smart”: Understanding the Security and Privacy Attitudes of Smart Home Users on Reddit},
  author={Li, Jingjie and Sun, Kaiwen and Huff, Brittany Skye and Bierley, Anna Marie and Kim, Younghyun and Schaub, Florian and Fawaz, Kassem},
  booktitle={2023 IEEE Symposium on Security and Privacy (SP)},
  pages={2850--2866},
  year={2023},
  organization={IEEE}
}

\appendix 
\newpage
\section{Availability} \label{sec:appendix_artifacts}
{\urlstyle{sf}
Study artifacts, including the interview consent form, screening survey, participants' full information, instructions used in design sessions, interactive prototype, and final codebooks, are linked at: {\color{blue}\url{https://osf.io/2w46p/overview?view_only=1bdc27b27b8e44b198638035995bb71a}}.
}

\section{Interview \& Design Session Script} \label{sec:appendix_interview}
\footnotesize
\subsection{Introduction \& Warm-Up Questions}
Hi, thank you for joining us today. My name is [NAME]. This is my colleague [NAME], who will be taking some notes in the background. In this interview, I will be asking you about your usage, experiences, and preferences for GenAI tools.

As mentioned in the consent form, we would like to audio-record this interview for analysis purposes. I will keep my video on throughout the interview process, but it is completely up to you if you would like to have your video on or off. May I confirm with you that you allow us to audio-record this interview? [Start the audio recording].

Great! Let’s get started with some warm-up questions first, to get to know your knowledge about GenAI tools. There are no right or wrong answers. We’re just interested in your opinions and experiences.

\begin{itemize}
    \item How would you define artificial intelligence, or AI, in your own (a few) words?
    \item How would you define Generative AI or GenAI, in your own (a few) words?
\end{itemize}

In this interview, when we refer to ‘GenAI tools,’ we mean both standalone GenAI apps, such as ChatGPT, as well as GenAI features built into other platforms, apps, or smart devices. For example, that could include voice assistants like Alexa+, AI-generated summaries on social media platforms, smart replies in messaging apps, or personalized recommendations in streaming services. 

Before we get started, do you have any questions or need any clarification about what would be counted as a GenAI tool?

\subsection{Consumer-Facing GenAI Tool Usage (RQ1)}
\begin{enumerate}
    \item What GenAI tools have you used so far?
    \item What are the reasons or purposes you have used these tools for?
    \item Have you ever used any tool that you no longer use, and why?
    \item \colorbox{mygray}{\textbf{[For participants using multiple GenAI tools]:}} Which of these GenAI tools do you use most often?
    \begin{enumerate}
        \item How long have you been using it, and how frequently do you use it?
        \item What is the main purpose for which you use this GenAI tool?
    \end{enumerate}
\end{enumerate}

\subsection{Pre-Adoption Considerations (RQ1)}
Now we would like to ask you some questions about your experiences and perceptions toward the GenAI tools before you adopted them. If your answers differ for some specific GenAI tools, please let me know.

\begin{enumerate}[start=5]
    \item What factors did you consider when deciding what GenAI tool to use?
    \item What information did you seek out, if any, to inform your decision to adopt these GenAI tools?
    \begin{enumerate}
        \item \colorbox{mygray}{\textbf{[If none]:}} If you are interested in adopting a GenAI tool now, what information would you be interested in looking for?
    \end{enumerate}
    \item Were there any other GenAI tools you’ve considered but decided not to use? Why did you decide not to use them/it?
\end{enumerate}

\subsection{Post-Adoption Considerations (RQ1)}
Now we would like to ask you some questions about your experiences and perceptions of the GenAI tools after you adopted them. Again, if your answers differ for some specific GenAI tools, please let us know.

\begin{enumerate}[start=8]
    \item On a scale of 0-10, 0 being not at all beneficial and 10 being very beneficial, how beneficial do you consider these GenAI tools to be, and what are the benefits if any?
    \item On a scale of 0-10, 0 being not at all harmful and 10 being very harmful, how harmful do you consider these GenAI tools to be, and what are the harms if any?
    \item \colorbox{mygray}{\textbf{[If S\&P concerns are not among the mentioned ones]:}} How concerned or comfortable are you with the privacy and security practices of these GenAI tools?
    \begin{enumerate}
        \item What are your security and privacy concerns?
        \item \colorbox{mygray}{\textbf{[If not at all concerned]:}} Are there reasons why you are not concerned?
    \end{enumerate}
    \item \colorbox{mygray}{\textbf{[If S\&P concerns are among the mentioned ones]:}} What are the security and privacy concerns you have?
    \item What steps, if any, have you taken to manage your concerns?
    \begin{enumerate}
        \item \colorbox{mygray}{\textbf{[If no steps have been taken]:}} What are the potential steps you could take to manage the concerns?
    \end{enumerate}
    \item What challenges do you think users face when managing these concerns?
\end{enumerate}

\subsection{Practices, Challenges, and Expectations Toward S\&P Information Seeking (RQ2\&RQ3)}

\begin{enumerate}[start=14]
    \item In practice, have you ever thought about privacy or security before adopting these tools or when interacting with them?
    \begin{enumerate}
        \item \colorbox{mygray}{\textbf{[If yes:]}} What about S\&P did you think about, and why were you interested in these topics?
        \begin{enumerate}
            \item What steps, if any, did you take to inform yourself about the privacy or security of these tools?
            \item Were you able to find all the information you were looking for? How?
            \item How confident were you about the trustworthiness of the information you found?
            \item Did you make any changes or take any actions based on the information you found?
            \item Was there any other S\&P information you wished to know?
            \item How easy or challenging did you find the process?
        \end{enumerate}
        \item \colorbox{mygray}{\textbf{[If not:]}} Could you please tell us why?
        \begin{enumerate}
            \item What aspects of S\&P might you be interested in and could affect your choice of GenAI tools? Why?
        \end{enumerate}
    \end{enumerate}
\end{enumerate}

\subsection{Design Session (RQ3)}
Thank you for your valuable insights! Next, we are doing some brainstorming about interfaces that can better help us manage our privacy and security concerns when using GenAI tools. To do this, we are doing a mini sketch session by using the whiteboard feature, as mentioned before the interview. [Launch the Zoom Whiteboard].

\noindent \colorbox{mygray}{\textbf{[Onboarding training (3 minutes):]}}
\begin{itemize}
    \item Ask the participant to read the design instructions (see Appendix~\ref{sec:appendix_artifacts}).
    \item Introduce the tools (pen, highlighter, eraser/undo, shapes, text).
\end{itemize}

\noindent \colorbox{mygray}{\textbf{[Task 1 (5-10 minutes):]}}
\begin{itemize}
    \item Scenario
    \begin{itemize}
        \item A consumer is considering signing up for a GenAI tool.
    \end{itemize}
    \item Goals
    \begin{itemize}
        \item Please create an interface that previews security and privacy information for the consumer prior to signing up.
        \item Elements you need to include:
        \begin{itemize}
            \item The structure/organization of the interface.
            \item The type(s) of security and privacy information that matter most to you.
            \item Visual elements (optional)
            \begin{itemize}
                \item Examples: buttons, windows
            \end{itemize}
            \item Interactive mechanisms (optional)
            \begin{itemize}
                \item Example: Pop-ups
            \end{itemize}
        \end{itemize}
    \end{itemize}
\end{itemize}

\noindent \colorbox{mygray}{\textbf{[Task 2 (5-10 minutes):]}}
\begin{itemize}
    \item Scenario
    \begin{itemize}
        \item A consumer has already created an account and is using the GenAI tool and they want to access some security and privacy information.
    \end{itemize}
    \item Goals
    \begin{itemize}
        \item Please create an interface for the consumer to access security and privacy information while using the tool.  If you would like the same design as you just did, please let us know.
        \item Elements you need to include:
        \begin{itemize}
            \item The structure/organization of the interface.
            \item The type(s) of security and privacy information that matter most to you.
            \item Visual elements (optional)
            \begin{itemize}
                \item Examples: buttons, windows
            \end{itemize}
            \item Interactive mechanisms (optional)
            \begin{itemize}
                \item Example: Pop-ups
            \end{itemize}
        \end{itemize}
    \end{itemize}
\end{itemize}

\noindent \colorbox{mygray}{\textbf{[Follow-up questions (5 minutes):]}}
\begin{itemize}
    \item Thank you! Whenever you're ready, could you walk me through your sketch? 
    \item Structure:
    \begin{itemize}
        \item Can you first give me a high-level overview of this design? What’s the overall layout or structure you had in mind?
    \end{itemize}
    \item Information
    \begin{itemize}
        \item What types of S\&P-related information are you including in this interface?
        \item Why did you choose to include these specific pieces of information?
    \end{itemize}
    \item Visual elements \& interactive mechanisms:
    \begin{itemize}
        \item I see you’ve labeled or highlighted a few things -- could you walk me through any visual elements or interactive features you’ve marked?
        \item What is the purpose of these features? What would they do or how would a user interact with them?
    \end{itemize}
    \item What problems are you trying to resolve through your design? What motivated you to create the design?
    \item What inspired your designs? Have you seen any other similar designs, interfaces, or mechanisms for security and privacy transparency of GenAI tools?
    \begin{itemize}
        \item \colorbox{mygray}{\textbf{[If yes:]}} How did you feel about them? What were the benefits and drawbacks?
    \end{itemize}
\end{itemize}

\subsection{Closing}
That brings us to the end of our session today! Thank you so much for your time and all the valuable insights you’ve provided! We are really glad that you decided to take part in our study. We will compensate you through Prolific in 5 to 7 days.

\section{Selected Original Sketches} \label{sec:appendix_original}

\begin{figure*}
    \centering
    \includegraphics[width=0.8\linewidth]{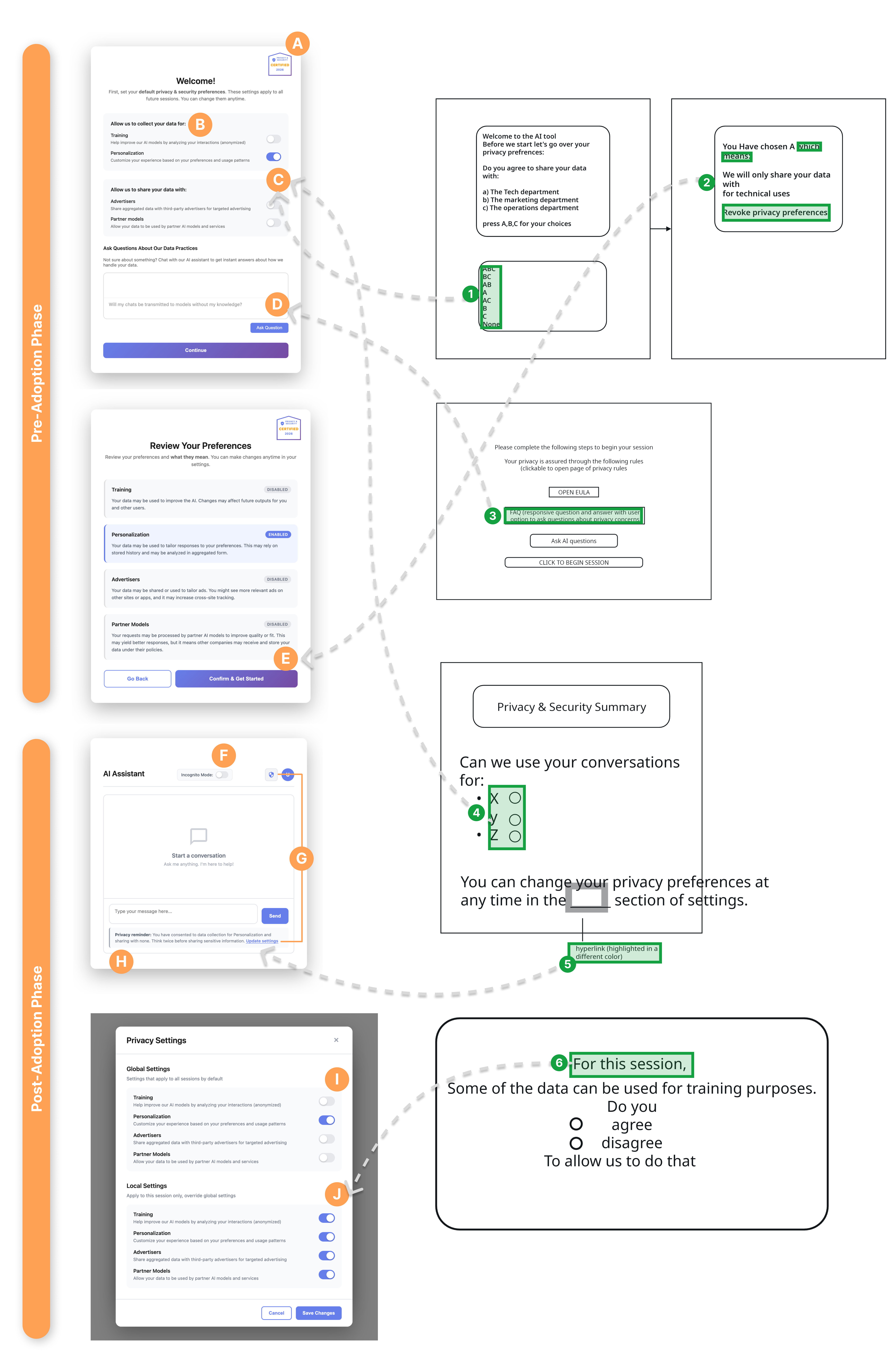}
    \caption{Examples showing traceability from participants' original sketch concepts to corresponding prototype features, including: granular permissions (\textbf{1, 4}), explanations and pre-commit editing of permissions (\textbf{2}), AI-enabled FAQs (\textbf{3}), clear visual navigation to privacy settings (\textbf{5}), and local, session-specific settings (\textbf{6}).}
    \label{fig:original}
\end{figure*}

\end{document}